\documentclass[a4paper, amsfonts, amssymb, amsmath, reprint, showkeys, nofootinbib, twoside]{revtex4-2}

\usepackage[english]{babel}
\usepackage[utf8]{inputenc}

\usepackage{amsthm}
\usepackage{mathtools}
\usepackage{physics}
\usepackage{xcolor}
\usepackage{graphicx}
\usepackage[left=13mm,right=13mm,top=25mm,bottom=25mm,columnsep=15pt]{geometry} 
\usepackage{adjustbox}
\usepackage{placeins}
\usepackage[T1]{fontenc}
\usepackage{lipsum}
\usepackage{csquotes}

\usepackage{fancyhdr} 

\fancyheadoffset{0cm}

\fancypagestyle{plain}{%
  \fancyhf{}%
  \fancyhead[R]{\thepage}%
}

\usepackage{xr-hyper} 
\makeatletter

\usepackage[pdftex, pdftitle={Article}, pdfauthor={Author}]{hyperref}
\usepackage{epstopdf}

\usepackage[left,mathlines]{lineno}

\begin{document}

\title{RF-Photonic Deep Learning Processor with \\ Shannon-Limited Data Movement}

\author{Ronald Davis III$^{1}$}
\author{Zaijun Chen$^{1,3}$}
\author{Ryan Hamerly$^{1,2}$}
\author{Dirk Englund$^{1}$}

\affiliation{$^{1}$Research Laboratory of Electronics, MIT, Cambridge, MA, 02139, USA}
\affiliation{$^{2}$NTT Research Inc., PHI Laboratories, 940 Stewart Drive, Sunnyvale, CA 94085, USA}
\affiliation{$^{3}$Ming Hsieh Department of Electrical and Computer Engineering, University of Southern California, Los Angeles, California 90089, USA}

\begin{abstract}

Edholm’s Law \cite{cherry2004edholm} predicts exponential growth in data rate and spectrum bandwidth for communications and is forecasted to remain true for the upcoming deployment of 6G.  Compounding this issue is the exponentially increasing demand for more deep neural network (DNN) compute \cite{xu2018scaling}, including DNNs for signal processing.  However, the slowing of Moore’s Law due to the physical limitations of transistor-based electronics means that completely new paradigms for computing will be required to meet these increasing demands for advanced communications.  Optical neural networks (ONNs) are promising DNN hardware accelerators with ultra-low latency and energy consumption. Yet state-of-the-art ONNs struggle with scalability and integrating linear matrix algebra  \cite{shen2017deep, zhu2022space, zhang2021optical, bagherian2018chip, xu2020optical, tait2017neuromorphic, feldmann2019all, bangari2019digital, xu202111, feldmann2021parallel, hamerly2019large, sludds2022delocalized, wang2022optical, farhat1985optical, kung1986optical, ohta1989new, zuo2019all, bernstein2022single, khoram2019nanophotonic, ashtiani2022chip, bandyopadhyay2022single} with optical nonlinear activations \cite{tait2019silicon, george2019neuromorphic, williamson2019reprogrammable, ashtiani2022chip, zuo2019all, jha2020reconfigurable, huang2020chip, crnjanski2021adaptive, feldmann2019all, basani2022all, bandyopadhyay2022single}. Here we introduce our multiplicative analog frequency transform optical neural network (MAFT-ONN) that encodes the data in the frequency domain, achieves matrix-vector products in a single shot using photoelectric multiplication \cite{hamerly2019large}, and uses a single electro-optic modulator for the nonlinear activation of all neurons in each layer.  We experimentally demonstrate the first hardware accelerator that computes fully-analog deep learning on raw RF signals, performing single-shot modulation classification with 85\% accuracy, where a `majority vote' multi-measurement scheme can boost the accuracy to 95\% within 5 consecutive measurements.  In addition, we experimentally demonstrate programmable frequency-domain finite impulse response (FIR) linear-time-invariant (LTI) operations, enabling a powerful combination of traditional and AI signal processing.  We also demonstrate the scalability of our architecture by computing nearly 4 million fully-analog multiplies-and-accumulates (MACs) for MNIST digit classification.  Our latency estimation model shows that due to the Shannon capacity-limited analog data movement, MAFT-ONN is hundreds of times faster than traditional RF receivers operating at their theoretical peak performance.

\end{abstract}

\maketitle

\section*{Introduction}
Artificial intelligence (AI) has been revolutionizing a broad range of fields, including RF signal processing.  In environments where the spectrum is congested with several users and many channels, hand-engineered systems are becoming increasingly infeasible.  Here, AI can be leveraged to process the increasingly complex spectral environment.

For RF signal processing, state-of-the-art AI approaches first digitize the IQ data and then either converts the signal into a $N \times 2$ IQ tensor \cite{lee2022exploiting, 8531759} or computes the spectrogram \cite{shen2021radio, REN2017225} (or other time-frequency transform) to convert the signal into an image.  The pre-processed signal is then inserted into a convolutional neural network (CNN) or other deep learning model for tasks like signal classification or device fingerprinting.

However, while digital processors can compute CNNs with high accuracy, these methods introduce significant latency that make real-time spectrum sensing impossible for digital processors. This is because state-of-the-art digital architectures require several steps to move the large volume of RF data to and from the compute.  The alternative is  to store the data for later offline analysis, but this is not feasible for time-sensitive tasks. In addition, these current approaches struggle to maintain high performance while keeping cost, size, weight, and power low.

Optical systems promise DNN acceleration by encoding, routing, and processing analog signals in optical fields, allowing for operation at the quantum-noise-limit with high bandwidth and low energy consumption. Optical neural network (ONN) schemes rely on (i) performing linear algebra intrinsically in the physics of optical components and/or (ii) in-line nonlinear transformations.  For (i), past approaches include  Mach-Zehnder interferometer (MZI) meshes \cite{shen2017deep, zhu2022space, zhang2021optical, bagherian2018chip, bandyopadhyay2022single}, on-chip micro-ring resonators (MRRs) \cite{xu2020optical, tait2017neuromorphic, feldmann2019all, bangari2019digital}, wavelength-division multiplexing (WDM) \cite{xu202111, feldmann2021parallel, sludds2022delocalized}, photoelectric multiplication \cite{hamerly2019large}, spatial light modulation \cite{wang2022optical, farhat1985optical, kung1986optical, ohta1989new, zuo2019all, bernstein2022single}, optical scattering \cite{khoram2019nanophotonic}, and optical attenuation \cite{ashtiani2022chip}.  For (ii), past approaches include optical-electrical-optical (OEO) elements \cite{tait2019silicon, george2019neuromorphic, williamson2019reprogrammable, ashtiani2022chip, bandyopadhyay2022single} and all-optical \cite{zuo2019all, jha2020reconfigurable, huang2020chip, crnjanski2021adaptive, feldmann2019all, basani2022all} approaches.  However, to fully take advantage of the potential ultra-low latency and energy consumption available in photonics, it is necessary to implement \textit{linear and nonlinear operations} together with minimal overhead.  Simultaneously achieving (i) and (ii) in a way that preserves high hardware scalability and performance has been an open challenge. 

\begin{figure*}[t!]
\centering
\includegraphics[width=\textwidth]{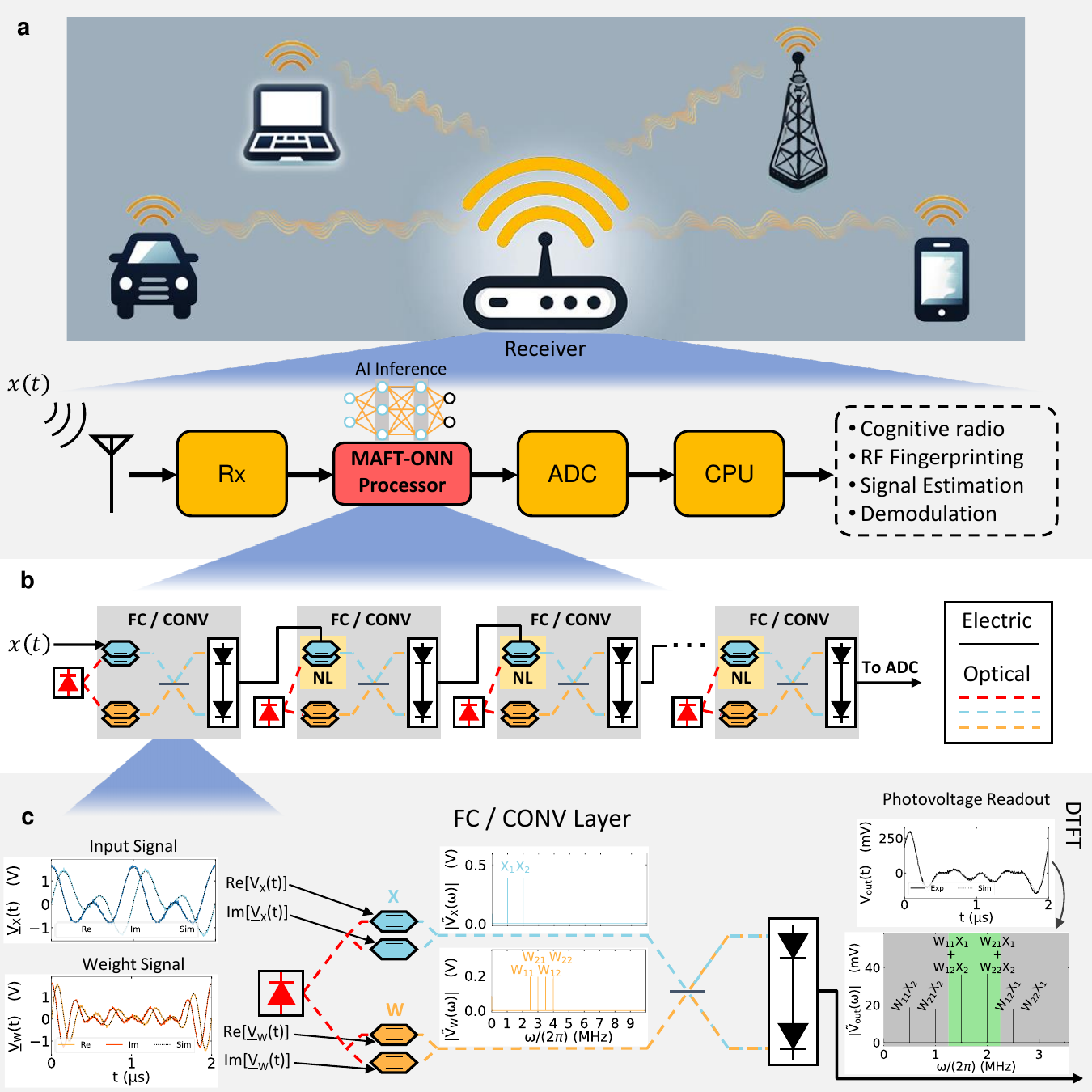}
\caption{An overview of the MAFT-ONN architecture.  (a)  The MAFT-ONN processor accelerates both traditional signal processing operations and AI inference for waveforms like radio waves.  The analog received waveform is fed into MAFT-ONN for fully-analog processing, after which the output may be read out digitally using an analog-to-digital converter (ADC) or fed into another analog system.  (b) An outline of the MAFT-ONN architecture.  Each photoelectric multiplication physically computes either a fully connected (FC) or a 1D convolution (CONV) layer.  The nonlinear activation (NL) for each layer physically corresponds to the nonlinear region of the following modulator.  The units can be cascaded to implement several DNN layers fully in analog with no digital overhead.  (c) A close-up of a single FC or CONV layer.  For a FC layer, the weight signal is programmed such that the photoelectric multiplication yields a matrix-vector product in the frequency domain, where the green region in the frequency domain is isolated with a filter.  For a CONV layer, all frequencies of the output signal are used.}
\label{fig:A}
\end{figure*}

Our multiplicative analog frequency transform optical neural network (MAFT-ONN) architecture simultaneously achieves (i) and (ii) for DNN inference with high scalability in both DNN size and layer depth.  We experimentally demonstrate the MAFT-ONN in a 3-layer DNN for inference of MNIST images and modulation classification.  In this architecture, we encode neuron values in the amplitude and phase of frequency modes, and `photoelectric multiplication' \cite{hamerly2019large} performs matrix-vector products in a single shot.  The nonlinear activation for each layer is achieved using the nonlinear region of an electro-optic modulator, thus enabling a scalable front-to-back photonic hardware accelerator for DNNs.  Figure \ref{fig:A}(a) contextualizes use-cases for the MAFT-ONN processor.

\section*{MAFT-ONN Architecture}

As illustrated in Figure \ref{fig:A}(b), a series of DNN layers corresponds to a cascading of photoelectric multiplications where each one computes either a fully connected (FC) or 1D convolution (CONV) layer.  The nonlinear activation (NL) for all neurons in a given layer is achieving by operating in the nonlinear regime of the following modulator.

Figure \ref{fig:A}(c) details an experimental example of a $2\times2$ matrix-vector product using the MAFT scheme.  In practice, the electrical voltage signal $V_X(t)$ is the incoming RF waveform and $V_W(t)$ is the weight signal generated by MAFT-ONN.   With an underbar denoting the complex representation of the signal, the terms $\text{Re} \left[ \underline{V}(t) \right ]$ and $\text{Im} \left[ \underline{V}(t) \right ]$ represent a voltage signal and its $90^\circ$ phase-shifted copy, the combination of which is required to achieve single-sideband suppressed carrier (SSB-SC) modulation. (See the Methods section for experimental details.)

The SSB-SC modulated input and weight signals are then photoelectrically multiplied to yield the output voltage signal:  $V_\text{out}(t) \propto \text{Im} \left[\underline{V}_X^*(t)\underline{V}_W(t)\right]$.

The output signal's Fourier Transform $\Tilde{V}_\text{out}(\omega)$ corresponds to the desired matrix multiplication performed in a single shot.  This matrix multiplication is achieved by appropriately programming the frequency content of the weight signal $V_W(t)$.  For an FC layer, the frequencies of $\Tilde{V}_\text{out}(\omega)$ that correspond to the matrix product are within the green region in Figure \ref{fig:A}(c) and are isolated using a bandpass filter.  For a CONV layer, all frequencies remain.  See Supplementary Section G for the generalized matrix multiplication algorithm.

\section*{Results}

\begin{figure*}[!t]
\includegraphics[width=\linewidth]{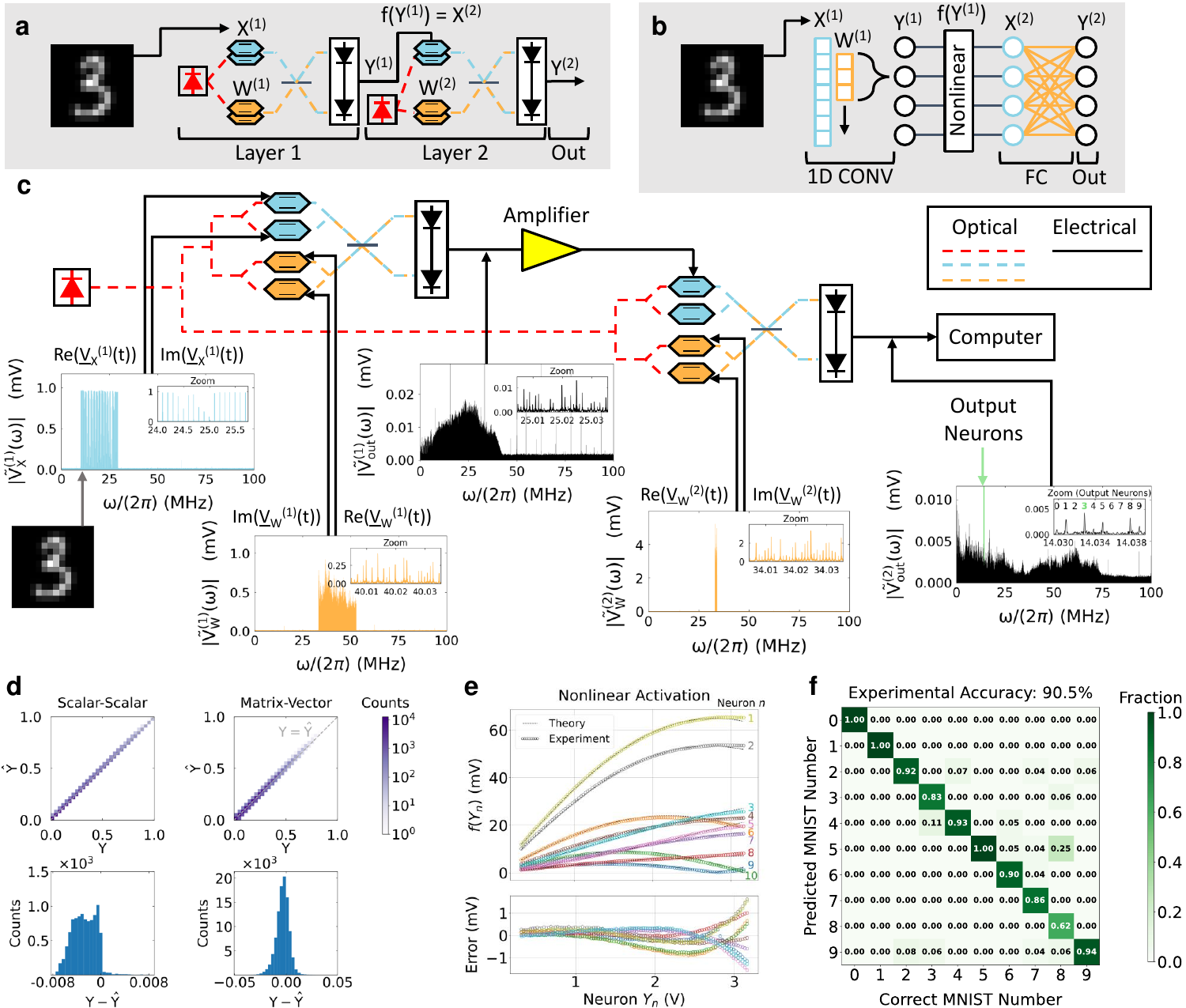}
\centering
\caption{The experimental demonstration of the MAFT-ONN.  (a)  An outline of the 3-layer MAFT-ONN.  (b)  The experimental DNN consists of two CONV layers with a nonlinear activation in the hidden layer.  (c)  A breakdown of the experimental inference of an $14\times14$ MNIST image as the input signal $V_{X}^{(1)}(t)$.  The two convolutions are achieved with the weight signals $V_{W}^{(1)}(t)$ and $V_{W}^{(2)}(t)$, respectively.  The nonlinear activation is achieved using an amplifier so that the signal $V_{\text{out}}^{(1)}(t)$ reaches the nonlinear region of the DPMZM.  As shown in the zoom plot of $V_{\text{out}}^{(2)}(t)$, the image is correctly classified.  (d)  In the upper half, 2D histograms compare the experimental output values $\hat{Y}$ to the expected curve fitted value $Y$.  In the lower half, 1D histograms plot the error $Y - \hat{Y}$.  The scalar-scalar plot contains 10,000 randomized $1 \times 1$ matrix products, yielding 9-bit precision compared to the curve fit.  The matrix-vector plot contains 10,000 randomized $10\times10$ matrix products (thus 100,000 values), yielding 8-bit precision.  (e)  An experimental characterization of the nonlinear activation function of an MZM.  We programmed $V^{(1)}_X(t)$ as a $10\times1$ input vector, and gradually increased its amplitude until it reached the nonlinear regime of the MZM.  We then curve fitted an analytical model to the experimental data.  (f)  A confusion matrix of the experimental 3-layer DNN over 200 $14\times14$ MNIST images, yielding an experimental accuracy of 90.5\%.}
\label{fig:exp}
\end{figure*}

\subsection*{MNIST Digit Inference}
We experimentally demonstrated a proof-of-concept of the MAFT-ONN architecture for a 3-layer DNN trained to classify MNIST digits. As shown in Figure \ref{fig:exp}(a), the DNN consists of two CONV layers with a nonlinear activation for the hidden layer.  The frequency-encoded input layer consisted of a flattened $14\times14$ MNIST image and thus contained 196 frequencies that represent the neurons.  This was convolved with a kernel of 19,600 frequencies to yield the hidden layer of 39,100 neurons.  After using a DPMZM as the nonlinear activation, the hidden layer was next convolved with a kernel of 1,000 frequencies to yield the output layer of 10 neurons, one output neuron for each of the MNIST digits.

The number of multiply-and-accumulates (MACs) computed per MNIST image inference is the sum of the MACs in two CONV layers.  The number of MACs computed for a 1D convolution between two vectors with lengths $N_1$ and $N_2$ is simply $N_1 N_2$.  This is because the number of MACs when the two vectors fully overlap (assuming $N_2>N_1$) is $\left(N_2 - N_1 + 1\right)N_1$ and the number of MACs when the convolution is at the edges is $2 \sum_{i=1}^{N_1-1}{i}=\left(N_1-1\right)N_1$.  Therefore the number of MACs experimentally computed per MNIST inference is:  $(14\cdot14) \cdot 19,600 + 1,000 \cdot 10 = 3,851,600 \ \text{MACs}$.  The first term is from convolving the flattened input image with the weights and the second term only counts the MACs used for the output neurons.

The 3-layer experimental DNN inferred 200 $14\times14$ MNIST images, where the digital DNN has an accuracy of 95.5\% and the experimental DNN has an accuracy of 90.5\%.  One contribution to the experimental inaccuracy are ripples found in the experimental nonlinear activation function, perhaps due to the path length difference of the interferometer (see Supplementary Section C for data on this).  A higher power low-noise amplified balanced photodetector would also increase the SNR of the signal going into the second layer.  Additionally, performing the DNN training in-situ on the hardware itself \cite{wright2022deep} could help better characterize it and increase the accuracy.  The confusion matrix of the experimental DNN is shown in Figure \ref{fig:exp}(f).

\begin{figure*}[!t]
\centering
\includegraphics[width=\linewidth]{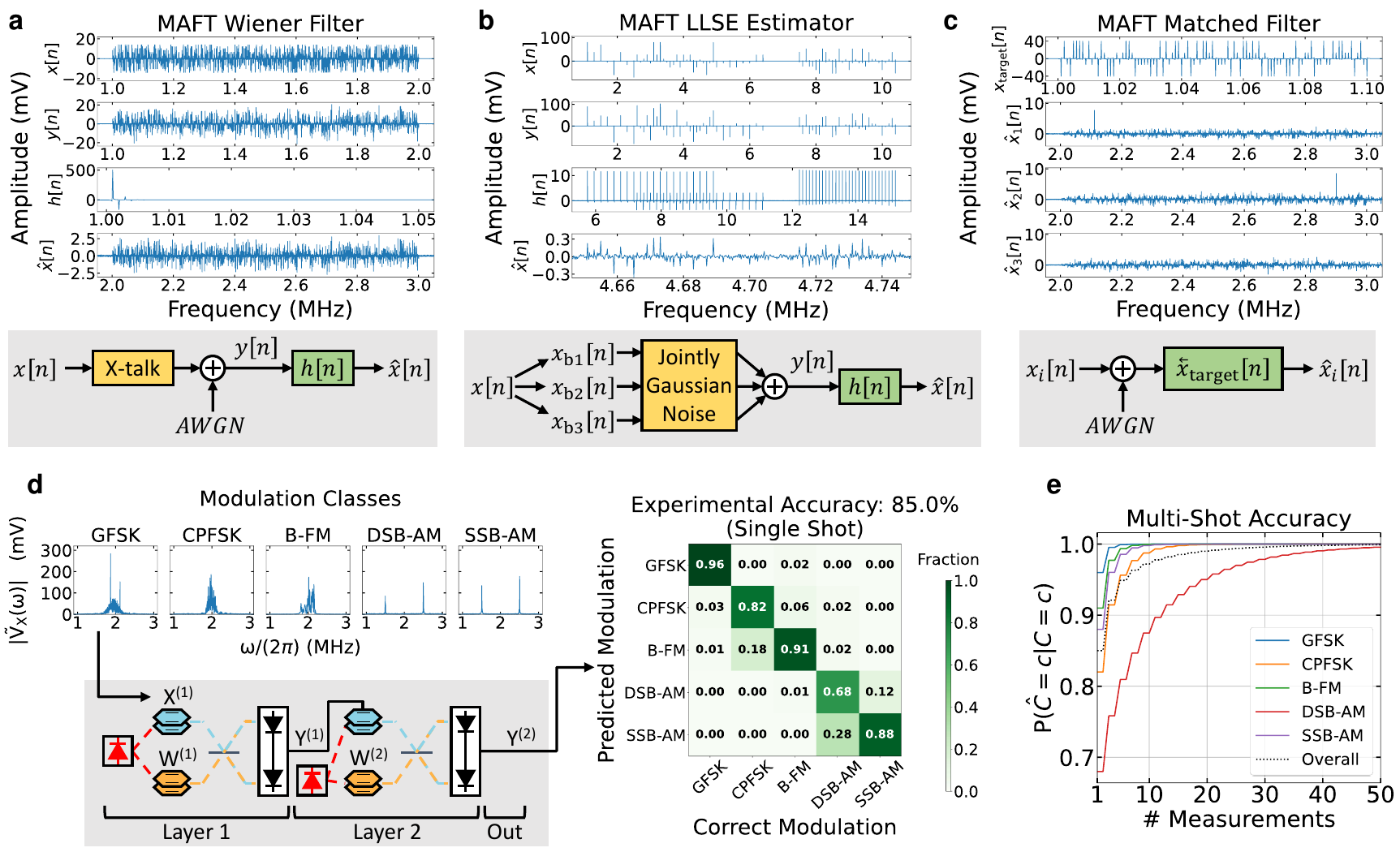}
\caption{Experimental demonstrations of the RF signal processing capabilities of MAFT-ONN.  The signal processing operations in (a)-(c) used the experimental setup in Figure \ref{fig:A}(c).  (a) The frequency-domain LTI framework was used to implement a Wiener filter that recovers a signal that suffered from frequency-crosstalk and AWGN.  (b) The matrix-vector product method from Figure \ref{fig:A}(c) was used to implement an LLSE estimator to recover a signal that suffered jointly Gaussian noise in various frequency bands.  (c) The frequency-domain LTI framework was used to scan the spectrum of a signal for a specific frequency signature.  (d) The 3-layer DNN hardware in Figure \ref{fig:exp}(c) was used to experimentally implement modulation classification on raw RF signals using MAFT-ONN.  For single-shot inference, the modulation classification experimentally achieved 85.0\% accuracy compared to the digital accuracy of 89.2\%.  (e)  With the experimental single-shot accuracy as the baseline, we show that a `majority vote' multi-measurement scheme will asymptotically improve the accuracy to 100\%.  With only 5 measurements, the accuracy improves to 95\%.}
\label{fig:rf}
\end{figure*}

\subsection*{LTI Signal Processing}

The MAFT architecture is capable of computing arbitrary finite impulse response (FIR) linear time invariant (LTI) operations in the frequency domain.  When interpreting the frequency modes as discrete LTI samples, the weight signal $V_W(t)$ can be programmed to implement arbitrary frequency-domain FIR LTI convolutions with the input signal $V_X(t)$.

Let the input voltage signal be $V_X(t)=\sum_{n=1}^{N}{X_n \cos(n \cdot \Delta \omega \cdot t)}$ and the LTI filter signal be  $V_W(t)=\sum_{r=1}^{R}{W_r \cos(r \cdot \Delta \omega \cdot t)}$.  Then the frequency-domain LTI interpretation is:  $x[n] \equiv X_n \rightarrow n \Delta \omega$ and $w[n] \equiv W_n \rightarrow n \Delta \omega$.  Hence the values of $x[n]$ and $w[n]$ are mapped to the frequency mode at $n \Delta \omega$.  With this interpretation, appropriately programming the SSB-SC conditions of the DPMZMs enables MAFT-ONN to compute convolutions that correspond to $y[n] = (x^* *w)[n] = \sum_{k=-\infty}^{\infty}{x^*[k]w[n-k]}$.  Therefore MAFT yields an LTI convolution for real-valued signals and a still useful LTI-like operation for complex-valued signals.  See Supplementary Section N for the mathematical derivation of the LTI framework from physics principles.

Figures \ref{fig:rf}(a)-(c) illustrate experimental results of various signal processing operations.

\noindent \textbf{FIR Wiener Filter:}  Let $V_X(t)$ be a frequency channelized signal with 1,000 frequency modes where the amplitude of each frequency encodes a symbol.  Here we assume a 4-bit encoding scheme where each symbol is equiprobable a priori.  After $V_X(t)$ is generated, crosstalk is introduced between each of the frequency modes and then additive white Gaussian noise (AWGN) is added to yield the distorted signal $V_Y(t)$.  The goal is to estimate the original signal $V_X(t)$ from the received signal $V_Y(t)$.

The frequency content of $V_X(t)$ and $V_Y(t)$ are modeled as jointly wide-sense stationary (WSS) random processes $x[n]$ and $y[n]$, respectively.  Given that the frequency crosstalk and AWGN are known, we calculate the covariance matrices to construct an FIR Wiener filter $h[n]$ of length 50, meaning that the previous 50 samples of $x[n]$ will be used to estimate the next sample.  The estimated signal $\hat{x}[n]$ is then computed by applying the Wiener filter to yield:  $\hat{x}[n] = (y*h)[n]$.

The effectiveness of the Wiener filter is evaluated using the mean squared error (MSE) between the original signal and the estimated signal.  The MSE is calculated over 10 experimental measurements of the instance of the random process.  The experimental MSE of the uncorrected signal $\left\| x - y \right\|^2$ has mean $39.46 \text{mV}^2$ and standard deviation $0.27 \text{mV}^2$, and the experimental MSE of the corrected signal $\left\| x - \hat{x} \right\|^2$  has mean $24.59 \text{mV}^2$ and standard deviation $0.14 \text{mV}^2$, yielding an average MSE improvement of 37.69\%.   This closely matches the theoretical MSE improvement of 38.78\%.  Figure \ref{fig:rf}(a) plots the one of the measurements for each signal.

According to Parseval's Theorem, estimating $V_X(t)$ in the frequency domain (opposed to the time domain) is effective since the energy of the error signal is the same in both the time and frequency domains.  Hence, given the appropriate context the MAFT frequency-domain estimation is a practical tool for signal processing scenarios.

\noindent \textbf{LLSE Estimator:}  The linear least squares estimator (LLSE) is more general than the Wiener filter and works on non-WSS random processes.  Let $V_X(t)$ be frequency channelized as in the previous example, except this time it is split up into three frequency bands each with different amplitude encoding characteristics.  In frequency-domain LTI notation $V_X(t)$ is broken into its three frequency bands as $x[n] = x_{\text{b}1}[n] + x_{\text{b}2}[n] + x_{\text{b}3}[n]$.

Each of the three bands goes through a combination of shared and individual sources of noise, which is modeled as jointly Gaussian noise, to yield the distorted signal $y[n]$.  Because each of the frequency bands have different encoding schemes and different sources of noise, $x[n]$ and $y[n]$ are no longer WSS and cannot be estimated using a Wiener filter.

Assuming we have knowledge of the jointly Gaussian noise, we constructed the optimal LLSE filter by calculating the appropriate covariance matrices.  We use the scheme to create the filter $h[n]$ that implements the LLSE estimator to produce $\hat{x}[n]$.

As before, 10 experiments of the instance of the were measured to calculate the experimental MSE.  The experimental uncorrected MSE $\left\| x - y \right\|^2$ has mean $479.58 \text{mV}^2$ with standard deviation $1.03 \text{mV}^2$, and the experimental corrected MSE $\left\| x - \hat{x} \right\|^2$ has mean $193.38 \text{mV}^2$ and standard deviation $22.98 \text{mV}^2$ for an average MSE improvement of 59.68\%.  This again is close to the theoretical MSE improvement of 63.17\%.  Figure \ref{fig:rf}(b) plots one of the measurements for each signal.

\noindent \textbf{Matched Filters:}  A useful benefit of MAFT is that a single photoelectric multiplication can be used to simultaneously scan the entire RF spectrum for a target signal.  Let $V_i(t)$ be a series of signals that we wish to scan for a specific frequency signature $\tilde{V}_\text{target}(\omega)$.  The frequency-domain LTI representations of $V_i(t)$ and $\tilde{V}_\text{target}(\omega)$ are $x_i[n]$ and $x_\text{target}[n]$, respectively.

Figure \ref{fig:rf}(c) shows three examples of applying the matched filter to various received signals.  For $x_1[n]$, the target $x_\text{target}[n]$ is located lower in the spectrum, and thus a peak appears there in the matched filter output.  For $x_2[n]$ the target is located higher in the spectrum so the peak appears there.  The signal $x_3[n]$ is a control that does not contain the target.

\subsection*{Modulation Classification}

Modulation classification entails identifying various schemes used to wirelessly transmit information, used in scenarios like cognitive radio.  Using the same experimental setup in Figure \ref{fig:exp}(c), the input activation signal $V_{X}^{(1)}(t)$ consisted of a frame from a synthetically generated waveform across five different types of modulation:  Gaussian frequency shift keying (GFSK), continuous phase frequency shift keying (CPFSK), broadcast frequency modulation (B-FM), double sideband amplitude modulation (DSB-AM), and single sideband amplitude modulation (SSB-AM) as illustrated by Figure \ref{fig:rf}(c).  The channel model includes fading due to Rician multipath, phase and frequency offsets due to clock offsets, timing drifts due to clock offsets, and AWGN such that the SNR is 30.  This 3-layer DNN experimentally achieved 85.0\% single-shot modulation classification accuracy over 500 input activation frames compared to the digital single-shot accuracy of 89.2\%.

Figure \ref{fig:rf}(e) illustrates the performance of a straightforward scheme to use multiple consecutive measurements to boost the accuracy.  When MAFT-ONN makes multiple inferences, the final inference can be determined with a `majority vote' scheme, where the final inference is simply the class with the most inferences.  The diagonal elements of the confusion matrix in Figure \ref{fig:rf}(d) yield $\text{P}( \hat{C} = c | C = c )$, which is the conditional probability that the inference $\hat{C}$ will be correct given the input class $C$.  The majority vote scheme assumes (i) that each measurement is statistically independent, (ii) the experimental single-shot accuracy represents the conditional probability for all incoming samples, (iii) the modulation class does not change in the duration of measurements, and (iv) only those five classes arrive at MAFT-ONN.  With this scheme, the conditional probability of correct detection quickly approaches 100\%, while reaching 95\% with only 5 measurements.

We note that the chosen modulation classes are inherently more oriented towards frequency-domain encoding compared to modulation schemes like quadrature amplitude modulation (QAM), pulse amplitude modulation (PAM), and phase shift keying (PSK).  In theory all of the information in even a purely phase modulation scheme like BPSK is still fully expressed in the frequency domain, albeit highly condensed around the carrier frequency.  As a proof-of-concept, we trained a digital twin of MAFT-ONN to distinguish between BPSK and QPSK with 90\% accuracy with a single layer.  This was enabled by appending an analog pre-processor before MAFT-ONN that applied an envelope detector and separately extracted the instantaneous frequency, thus spreading the phase information more into the frequency domain.

\section*{Discussion}

\begin{figure*}[!t]
\centering
\includegraphics[width=\linewidth]{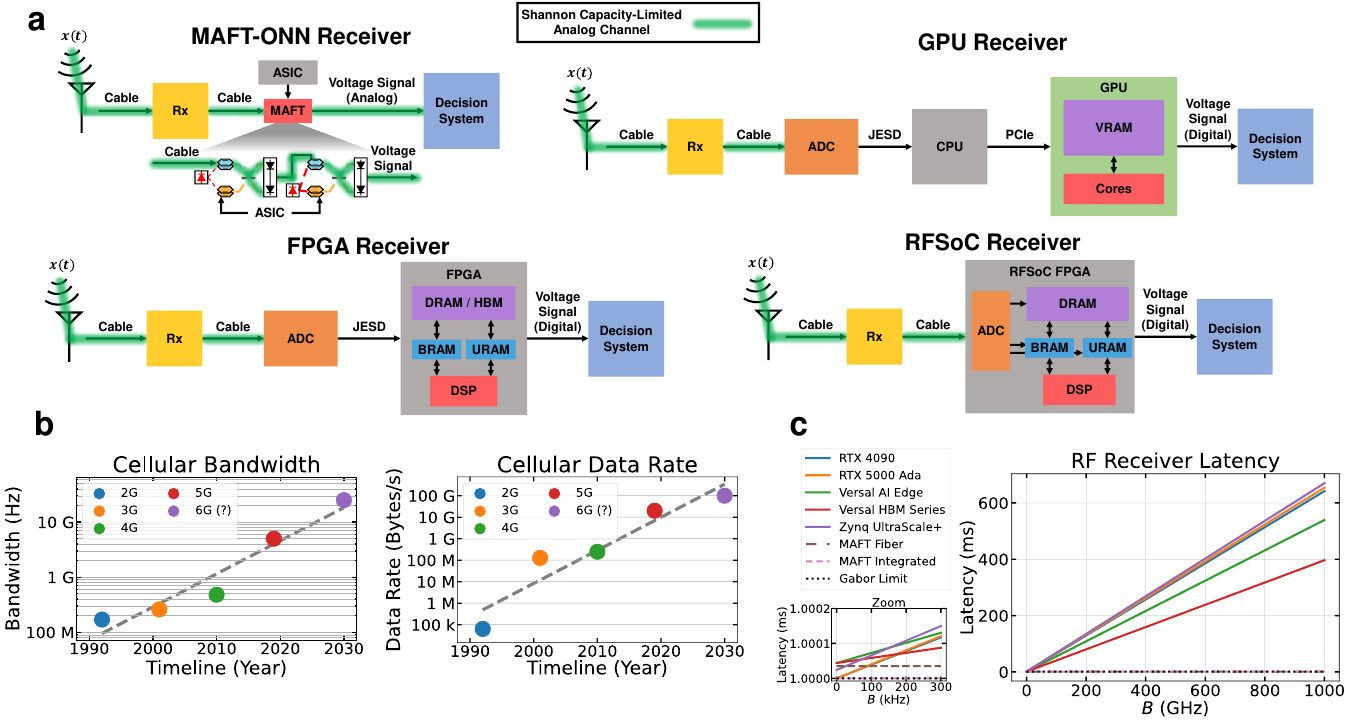}
\caption{System-level latency analysis comparing the MAFT-ONN architecture to state-of-the-art digital architectures.  (a) Modeling various high-performance computing architectures in the context of receiving an RF signal $x(t)$, implementing a filter on it, and then sending it to a `decision system' that performs some operation based on the inferred qualities of the processed $x(t)$.  The green highlights show that the analog information `flows' into and through MAFT-ONN, and thus the amount of information moving through the MAFT-ONN processor is limited by the Shannon capacity of the analog channels feeding it information (in the absence of non-ideal factors like multi-path fading and interference).  (b) An example of Edholm's Law applied to mobile communications.  This shows that wireless mobile bandwidth and data rates have been and will likely continue to exponentially grow.  (c)  Using the models in (a), the latency for processing $x(t)$ with increasing bandwidth $B$ is estimated for various high-performance processors.  We assume that the digital processors are performing at peak specifications with 100\% processor and memory utilization in ideal conditions.  For simplicity we assume that all FPGA components are running at 500 MHz.  Because MAFT-ONN avoids the bottleneck of digital data movement, its overall latency improves by two orders of magnitude compared to traditional receivers.}
\label{fig:lat}
\end{figure*}

To our knowledge, the MAFT architecture is the first hardware accelerator that performs DNN inference on raw communications signals.  Our several experimental demonstrations show that MAFT is flexible enough to implement a powerful combination of fully-analog frequency-domain LTI processing and DNN inference to enable new high-performance signal processing capabilities with the benefits of the low power consumption, cost, size, and weight of optical systems.  The power consumption of optical systems with different architectures but similar components has already been extensively studied \cite{hamerly2019large, bandyopadhyay2022single}, so Supplementary Section O analyzes the communication link gain of MAFT-ONN.

\subsection*{Computational Throughput}

The throughput $T$ measures the number of multiply-and-accumulates (MACs) computed within a given time.

\begin{linenomath}
\begin{align*}
T &= \frac{\text{\#\ MACs}}{\text{latency}}. \nonumber
\end{align*}
\end{linenomath}

The number of MACs computed in an FC layer with $N$ input neurons and $R$ output neurons is $N\cdot R$ MACs.  A CONV layer can be interpreted as keeping the spurious frequencies in an FC layer; thus the number of MACs in a CONV layer with $N$ input neurons and $N \cdot R$ weight neurons is $N^2 \cdot R$ MACs.  The time it takes to read out the output signal is the Gabor limit, which is $\frac{1}{\min(\Delta f, f_0)}$, where $\Delta f$ is the smallest frequency spacing of the output signal and $f_0$ is the lowest neuron frequency of the output signal.  (Note that $2 \pi f = \omega$, where $\omega$ is the angular frequency from the previous sections.)

Supplementary Sections H and I derive the theoretical throughput under these conditions, showing that the throughput for an FC layer asymptotically approaches the available bandwidth $B$ for modulating signals: $T_\text{FC} \approx B$, and similiary for the throughput for a CONV layer: $T_\text{CONV} \approx N \cdot B$.

The bandwidth $B$ limiting the throughput is not the RF bandwidth of the electrical components, but the available optical bandwidth, for example the 20 THz among the S, C, and L telecommunications bands.  See Supplementary Section M for an analysis of scalable MAFT-ONN architectures that take advantage of the full optical bandwidth and multiplexing.

Therefore, the combination of using the full optical bandwidth (on the order of terahertz) and spatial multiplexing (on the order of a hundred)\cite{streshinsky2014silicon} immediately yields a straightforward path to reaching peta-operation per second scale throughputs using MAFT-ONN with current technology.  Thus, the MAFT-ONN is competitive with electronic counterparts like the Google TPUv3 that has a throughput greater than 400 tera-operations per second \cite{kumar2019scale}.

The throughput for a given bandwidth $B$ is doubled by using complex-valued operations.  The encoding scheme for computing complex-valued matrix products with MAFT-ONN is derived in Supplementary Section L.

\subsection*{System-Level Latency}

Figure \ref{fig:lat}(a) illustrates how state-of-the-art digital architectures require several steps to move data to and from the compute (where the compute elements are the red sub-blocks).  This bottleneck is only being accentuated by Edholm's Law, which predicts that the amount bandwidth and data rates for communications will continue to exponentially rise.  Figure \ref{fig:lat}(b) illustrates an example of Edholm's Law for mobile communications, plotting the year each mobile network generation was introduced and its maximum bandwidth and data rate.  We compare the theoretical system-level latency for various RF receiver architectures in Figure \ref{fig:lat}(c), where system-level latency is defined as the duration of time between the signal first arriving at the antenna to the time the decision making system can extract the desired information from the output voltage signal.

In this scenario we assume that the incoming signal $x(t)$ has bandwidth $B$ and that the computations for $x(t)$ require frequency resolution $\Delta f$.  For simplicity we assume that the pre-selector circuit labeled `Rx' contributes negligible latency and that the ADC speed exactly matches the required Nyquist sampling rate for $B$.  Together, $B$ and $\Delta f$ determine the number of points $N=2 \lceil \frac{B}{\Delta f} \rceil$ required to digitally capture $x(t)$.  Therefore as $B$ increases, the amount of digital data that must be moved to/from memory also increases, thereby raising the digital latency.

We used the models in Figure \ref{fig:lat}(a) to estimate the theoretical latencies for the digital processors in Figure \ref{fig:lat}(c).  We assumed that all processors and data lanes are operating:  at their theoretical peak speeds, with 100\% processor and memory utilization, with no processing or memory management overhead, with no data rate degradation from high clock speeds, with no throttling due to overheating, and limitless DRAM/HBM in which to store RF data.  In practice these processors may experience over an order of magnitude slower latency than their theoretical optimal performance. The FGPAs will likely incur less overhead than the GPUs since the FGPAs trade off having less management overhead by storing less memory.

Per the scenario in Figure \ref{fig:lat}(a), after $x(t)$ is processed into an output voltage signal, it reaches a `decision system' that performs the appropriate operation based on the output signal's qualities.  For the traditional receiver architectures, the output voltage signal is digital (this contribution to the digital latency is neglected).  For MAFT, the output voltage signal is an analog waveform, so the decision system extracts the desired information from the waveform to perform its operation.  MAFT's decision system may or may not use an ADC, for example a non-ADC decision system may consist of a bank of bandpass filters followed by an analog power selector circuit.  Regardless of whether an ADC is used, the latency of MAFT's decision system is fundamentally dominated by the Gabor limit, which is included in MAFT's latency calculation.

As the Zoom subplot in Figure \ref{fig:lat}(c) illustrates, some digital processors theoretically have superior latency at low bandwidths, but even then quickly get overtaken by MAFT as the bandwidth increases.  At 15 GHz MAFT’s latency ranges from 7 to 11 times faster than digital architectures.  As the input signal bandwidth increases, so does MAFT's latency benefit.  For example, when considering scenarios like 5G multiple-input multiple-output (MIMO) where there are massive volumes of RF data streaming into the receiver, the combined instantaneous bandwidth is much larger than that of a single antenna receiver.  Figure \ref{fig:lat}(c) illustrates that at 1,000 GHz, MAFT's latency improvements range from 400 to 670 times faster than digital architectures.

The latency benefit from MAFT is physically attributed to the fast data movement to the compute.  In ideal communications scenarios, the Shannon capacity of the analog channel connecting the transmitter to MAFT is the limiting factor for moving the transmitted information to the signal processing computations.  The green highlighted paths in Figure \ref{fig:lat}(a) illustrate that the analog data `flows' into and through MAFT, and thus the amount of information that is processed by MAFT is determined by the capacity of the channel that feeds the information to MAFT.  And as discussed for the throughput, there is more than enough optical bandwidth to accommodate the channel capacity of incoming RF waveforms.

The latency per photoelectric multiplication of the MAFT-ONN experiment was measured to be 60 ns, which matches the expected latency for the propagation through fiber and photodetector bandwidth.  See Supplementary Section P for a breakdown of our latency estimation models.

\section*{Conclusion}

We introduced and demonstrated our MAFT analog computing scheme and MAFT-ONN hardware accelerator for scalable, fully-analog DNN acceleration that uses frequency-encoded neurons for convolutional and fully connected layers.  The MAFT-ONN architecture yields much flexibility for running various types and sizes of DNNs without changing the hardware.

This architecture is also the first DNN hardware accelerator that directly performs inference on raw RF signals, offering faster latency than FPGAs while simultaneously benefiting from the cost, size, weight, and power consumption of optics.  The combination of frequency-domain LTI signal processing and trainable weight signals enables a flexible, high-performance computing paradigm.  In addition, when using the full optical bandwidth and spatial multiplexing the throughput of this processor is also competitive with other state-of-the-art DNN hardware accelerators.

Future work includes increasing the scale of MAFT-ONN using wavelength-division and spatial multiplexing.

\section*{Methods}

\subsection*{2 $\times$ 2 Matrix Multiplication Experiment}

For the example experiment in Figure \ref{fig:A}(c) that details a $2 \times 2$ matrix multiplication, $V_X(t)$ and $V_W(t)$ were generated by an arbitrary waveform generator (AWG) and then sent to dual-parallel Mach-Zhender modulators (DPMZMs).  These DPMZMs perform single-sideband suppressed carrier (SSB-SC) modulation of the signals, without which the modulated signals would be dual-sideband and cancel each other out after the photoelectric multiplication.

To SSB-SC modulate a signal using a DPMZM, one copy of the signal is sent to one of the arms of the DPMZM while a $90^\circ$ phase-shifted copy is sent to the other arm.  Thus, let an underbar indicate an analytical Hilbert transform, yielding $\underline{V}_X(t) = \text{H}_a \left[ V_X(t) \right]$.  Then $\text{Re} \left[ \underline{V}_X(t) \right ] = V_X(t)$ is the original signal, and $\text{Im} \left[ \underline{V}_X(t) \right ]$ is the $90^\circ$ phase-shifted copy.  Although the $90^\circ$ phase-shifted was generated using an AWG in this experiment, in practice this phase shift can be achieved using commercial passive RF phase shifters.

\subsection*{3-Layer MNIST DNN}
We experimentally demonstrated the 3-layer DNNs using the MAFT-ONN scheme with the apparatus shown in Figure \ref{fig:exp}(c). An AWG generates $V_{X}^{(1)}(t)$, $V_{W}^{(1)}(t)$, and $V_{W}^{(2)}(t)$, all of which are modulated into the optical domain using DPMZMs.  The photoelectric multiplication of $V_{X}^{(1)}(t)$ and $V_{W}^{(1)}(t)$ yielded $V_{\text{out}}^{(1)}(t)$ after the first layer.  We then amplify $V_\text{out}^{(1)}(t)$ to reach the nonlinear regime of the DPMZM in the second layer, after which it became the input signal to the hidden layer.  For convenience, we only used one sub-modulator of the DPMZM for this signal, thus modulating $V_\text{out}^{(1)}(t)$ in the dual-sideband suppressed carrier (DSB-SC) mode.  For the hidden layer, we programmed $V_{W}^{(2)}(t)$ as the next CONV kernel.  The multiplication of the DSB-SC modulated $V_\text{out}^{(1)}(t)$ and the SSB-SC modulated $V_{W}^{(2)}(t)$ results in a copy of $V_{\text{out}}^{(2)}(t)$ appearing further up in the spectrum, as can be seen in the plot of $Y^{(2)}$ in Figure \ref{fig:exp}(c).  Finally, the analog output of the second layer $V_{\text{out}}^{(2)}(t)$ was sampled digitally and Fourier transformed.  See Supplementary Section A for more details on the 3-layer DNN experimental setup.

The input activations are downsampled $14\times14$ MNIST images that are represented by the frequency-encoded signal $V_{X}^{(1)}(t)$ containing 196 frequencies spaced at 100 kHz.  The input activation is convolved by a weight kernel $V_{W}^{(1)}(t)$ containing 19,600 frequencies spaced at 1 kHz to yield the signal of the hidden layer, $V_{\text{out}}^{(1)}(t)$. Note that $V_{W}^{(1)}(t)$ was programmed as for an FC layer, but we chose to keep the `spurious' frequencies thus effectively making it a CONV operation.  Next, the hidden layer $V_{\text{out}}^{(1)}(t)$, is multiplied by the second layer weight signal $V_{W}^{(2)}(t)$ that contains 1,000 frequencies spaced at 1 kHz to yield the output signal $V_{\text{out}}^{(2)}(t)$.

We implemented the one-hot vector that represents the output MNIST values by randomly selecting a set of 10 adjacent frequencies among the frequencies of $V_{\text{out}}^{(2)}(t)$ to demonstrate the flexibility of our scheme.  (See Supplementary Section K for a performance analysis of using sparse FC layers for DNN training.)  The 10 output neuron frequencies were randomly chosen to be 14.03 MHz to 14.039 MHz, with 1 kHz spacing.  The zoom of the plot of $V_{\text{out}}^{(2)}(t)$ in Figure \ref{fig:exp}(c) shows the mapping of the neuron frequencies to the MNIST digits, where the digit is classified by the frequency mode with that largest magnitude.  Hence the final readout only considers the absolute value of the frequencies modes, making the one-hot vector insensitive to phase.

For the MNIST classification we programmed real-valued positive and negative neuron values into both the input vectors and weight matrices.  Negative neuron values are physically represented by a $\pi$ phase shift in that particular frequency mode, allowing for analog matrix algebra with negative numbers.

An analytic model of the hardware was used to train the DNN offline, similar to the nonlinear characterization in the previous section.  The offline training produced a set of weight matrices that were then encoded into the RF signals used for the experimental inferences.  See Supplementary Sections B, C and E for details on the offline DNN characterization and training.

\subsection*{3-Layer Modulation Classification DNN}
The same hardware setup used for the MNIST classification in Figure \ref{fig:exp}(c) was also used to implement the 3-layer DNN for modulation classification, so we reuse the mathematical notation.  The input activation waveforms were generated using MATLAB \cite{MathWorksModulation}.  Each waveform had a random time delay applied in order to model receiving the signals at unknown times.  The power of all the frames was normalized at baseband and then upconverted to 1 MHz as shown in Figure \ref{fig:rf}(c).

The first DNN layer was CONV with the DPMZM biases configured such that $V_{W}^{(1)}(t)$ functioned as a trainable LTI filter.  Following the first CONV layer was the DPMZM nonlinearity.  Then, as portrayed in Figure \ref{fig:exp}(c), the second-layer weight signal $V_{W}^{(2)}(t)$ was DSB-SC modulated and implemented another CONV layer.  The LTI weight signal $V_{W}^{(1)}(t)$ consisted of 50 frequency modes spaced at ~9 kHz and the second-layer weight signal $V_{W}^{(2)}(t)$ consisted of 6,143 frequency modes spaced at ~200 Hz.  Both $V_{W}^{(1)}(t)$ and $V_{W}^{(2)}(t)$ were modeled as complex signals, so the number of trainable parameters was double the number of frequency modes (one for magnitude and phase).

For the one-hot vector readout of $V_{\text{out}}^{(2)}(t)$, instead of choosing adjacent frequencies like in the MNIST classification, we chose frequencies that were evenly spread out across the spectrum as we found that this allowed the weights to train most effectively.  And as with our MNIST inference the absolute value of the frequencies were measured so that the frequency mode with the largest magnitude was considered the DNN inference.

\section*{Acknowledgements}

\subsection*{Funding}

This work was funded by the Army Research Laboratory Electronic Warfare Branch (Stephen Freeman, Chief), the U.S. Army (W911NF-18-2-0048, W911NF2120099, W911NF-17-1-0527), the U.S. Air Force (FA9550-16-1-0391), MIT Lincoln Laboratory (PO\#7000442717), research collaboration agreements with Nippon Telegraph and Telephone (NTT), and the National Science Foundation (NSF) (81350-Z3438201).

\subsection*{Contributions}
D.E. and R.D. conceived the idea for the architecture.  D.E. conceived the idea for programming neurons in time-frequency space and the method of using a single balanced photodetector for the photoelectric multiplication to compute the matrix-vector product in time-frequency space.  R.D. conceived the MAFT frequency-encoding scheme that allows for single-shot photoelectric multiplication in the frequency domain by appropriately programming the inputs and weights and the method of using SSB-SC modulated signals for the MAFT-ONN hardware architecture.  R.D. developed the theoretical performance metrics, conducted the hardware simulations, and planned and executed the experiments.  Z.C. aided with planning and executing the hardware setup, characterization, debugging, and comparison with the theory.  R.D. finalized and analyzed the experimental results.  R.H. introduced the idea of using a discrete cosine transform to efficiently model the MZM nonlinearity, and R.D. conceived and created the offline physics-based DNN training algorithm.  R.D. conceived and developed the LTI signal processing framework, the latency estimation model, modulation classification, and performed the various signal processing experiments.  R.D. wrote the manuscript with contributions from all authors.  D.E. supervised the project.

Sampson Wilcox (MIT) aided in designing Figure 1.

\section*{Competing Interests}

The authors R.D. and D.E. disclose that they are inventors on pending patents US Application No. 63/315,403; US 63/638,172; and US 63/638,197 where MIT is the patent applicant.

\section*{Data and Materials Availability}

The data from this work will be made available upon reasonable request.

\bibliography{sample} 

\end{document}


\title{Supplementary Materials: RF-Photonic Deep Learning Processor with \\ Shannon-Limited Data Movement}

\author{Ronald Davis III$^{1}$}
\author{Zaijun Chen$^{1,3}$}
\author{Ryan Hamerly$^{1,2}$}
\author{Dirk Englund$^{1}$}

\affiliation{$^{1}$Research Laboratory of Electronics, MIT, Cambridge, MA, 02139, USA}
\affiliation{$^{2}$NTT Research Inc., PHI Laboratories, 940 Stewart Drive, Sunnyvale, CA 94085, USA}
\affiliation{$^{3}$Ming Hsieh Department of Electrical and Computer Engineering, University of Southern California, Los Angeles, California 90089, USA}

\maketitle

\tableofcontents

\cleardoublepage

\section*{Experiment}
\subsection{3-Layer DNN Hardware}
The experimental setup for the 3-layer DNN inference is illustrated in the main text.  All of the optics is done in-fiber using commercial components.  A total of four DPMZMs are used for the 3-layer experiment.  As illustrated in the main text, two pairs of DPMZMs implement the two matrix-vector photoelectric multiplications.

The first set of DPMZMs is from Exail, where all the fibers both before and after the modulator are polarization-maintaining (PM) with FC/APC leads.  The second set of DPMZMs is from Thorlabs, where all the fibers before the modulator are PM, but all the fibers after are single-mode (SM) with FC/PC leads.  Thus, we demonstrate that this architecture works with both types of fiber.

Since every DPMZM has 3 DC biases (one for each sub-MZM and then a phase bias for interfering the sub-MZMs together), each of the DPMZMs was controlled using Exail DPMZM bias controllers, which biased all the DPMZMs for SSB-SC modulation.  The bias controller does this by biasing each of the sub-MZMs at their minimum point, and then biasing the phase.  Note that the minimum intensity bias point of an MZM is usually highly nonlinear when using a single photodetector.  This is because for an electric field $E(t)$, the output of the signal photodetector is proportional to $\left|E(t) \right|^2$, the intensity.  This means that for a bias $\phi_b$, the single photodetector output will yield a bias curve of $\sin{\left(\phi_b\right)}^2=\frac{1}{2}\left(1 - \cos{\left(2 \phi_b\right)} \right)$.  However, the output of a balanced photodetector is proportional to not the intensity, but the electric field $\left|E(t) \right|$, yielding a bias curve of $\sin{\left(\phi_b\right)}$.  Thus, for a value of $\phi_b = 0$, we see that this is the minimum bias point of the intensity, but is the positive quadrature bias point of the electric field.  This conveniently allows us to compute linear matrix operations while operating in the SSB-SC mode.

After setting the bias points for all the modulators, the bias would remain stable for $>24$ hours without active control.  The Thorlabs DPMZMs require DC-blocks at the RF inputs to stabilize the bias points.

\begin{figure*}[t]
\centering
\includegraphics[width=\textwidth]{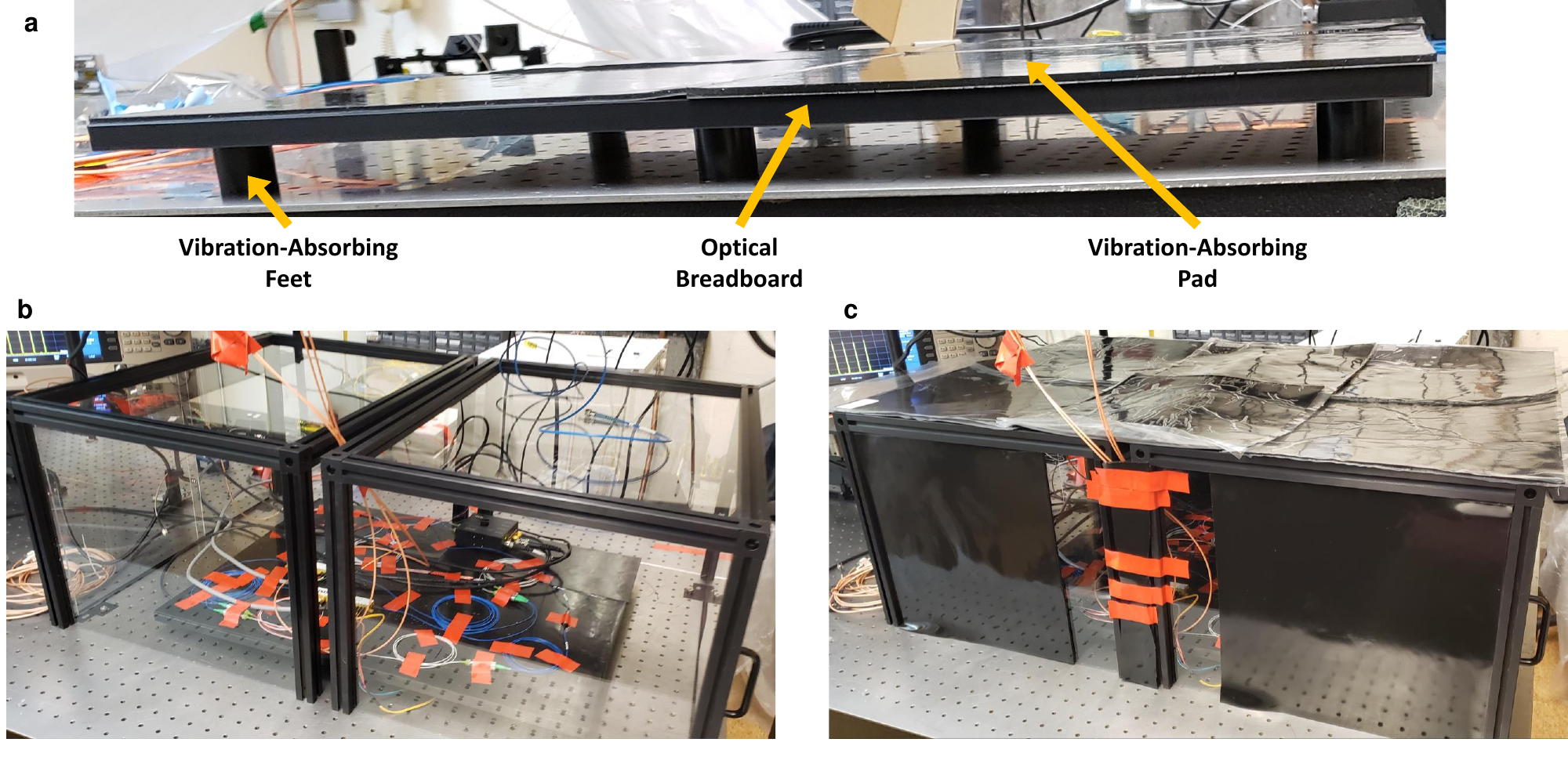}
\caption{The stabilization of the fiber interferometer that implements the first photoelectric multiplication in the 3-layer DNN experiment.  The whole setup is on a single optical table that does not have air stabilization.  (a)  Each interferometer has its own optical breadboard, supported by vibration-absorbing pads.  (b)  Each interferometer was boxed to reduce temperature and air current fluctuations.  (c)  More padding was placed on the box to reduce vibrations.}
\label{fig:stable}
\end{figure*}

Figure \ref{fig:stable} shows how the interferometers are stabilized.  The entire experiment is on an optical table that does not have air stabilization.  Each interferometer is placed on an optical breadboard that is lifted by vibration-absorbing feet.  The surface of each optical breadboard is also covered with vibration-abosrbing material on which the fibers and components are placed.  The fibers are also taped down to minimize movement.  Both interferometers are boxed to reduce temperature and air current fluctuations.

We use one 1550nm external cavity single-frequency laser with a temperature and power controller.  The laser is split into four paths; one for each of the DPMZMs.  Although the laser is pig-tailed with a PM fiber at the output, we observed the best stability when using fiber polarizers after the laser and before the DPMZMs.

The frequency-encoded RF signals were generated using two Keysight M3202A AWGs.  Each DPMZM requires two RF inputs to achieve SSB-SC modulation; one copy of the RF signal at one sub-modulator, and other copy of the RF signal with a  $90^{\circ}$ phase shift at the other modulator.  Thus, two AWG channels were used for each DPMZM.  Since one of the DPMZMs is driven by the output of the first layer, we used 6 AWG channels total for the 3-layer inference experiment (not including using one channel to ground the unused sub-MZM on the DPMZM operating in the DSB-SC mode).

For accurate results, the triggers of the AWG signals must be synchronized.  This is because any time delay will cause a gradient in the phases between all the frequencies, thus changing the behavior of the photoelectric multiplication.  As explained in the main text, we measured a 60 ns time delay between the first layer and the second layer, so we programmed the weight signal for the second layer weight matrix with a corresponding time delay for MNIST inference.  For the modulation classification we trained on signals with randomized time delays anyways so we didn't adjust the second layer weight signal timing.

We used a spectrum analyzer to read the amplitude of the output frequencies for the MNIST inference and an oscilloscope plus taking the absolute value of the Fast-Fourier Transform (FFT) for the modulation classification.

\subsection{Linear Curve Fit Characterizations}

Note that the following mathematical analysis uses the notation from Figure 3 in the main text.  All references to equations and figures to the main text will be explicitly stated.  All the linear curve fit experiments use real values that are programmed to be positive or negative using a $\pi$ phase shift.

\noindent \textbf{Mathematical Analysis:}  To test linear matrix-vector multiplication, we measure the photovoltage response $V_{\text{out}}^{(1)}(t)$ using a  spectrum analyzer that scans the relevant part of the bandwidth to extract $V_{Y}^{(1)}(t)$.  Here, the input laser is modulated by $V_{X}^{(1)}(t)$ and $V_{W}^{(1)}(t)$ via DPMZMs in the linear regime.  We repeat this multiplication over randomized values of $X^{(1)}$ and $W^{(1)}$ to obtain the full set of characterization data.

To measure the accuracy of the matrix products, we use a theoretical model to compare with the experiment.  From (main text) Equation 4, the result of linearly modulating the input vector is:

\begin{linenomath}
\begin{align}
f(V_{X}^{(1)}(t)) &= \chi_0 + \chi_1 e^{i \omega_{\text{LD}} t} \cdot \text{H}_a \left[ \sin \left(\chi_2 V_{X}^{(1)}(t) + \chi_3 \right) \right] \nonumber \\
& \approx \chi_1 \chi_2 e^{i \omega_{\text{LD}} t} \cdot \text{H}_a \left[V_{X}^{(1)}(t) \right] \nonumber \\
& \approx \chi_1 \chi_2 E^{(j)}_X(t)  \nonumber
\end{align}
\end{linenomath}

\noindent where we assume that $\chi_0 = \chi_3 = 0$ in the linear regime.  Similarly, the linear modulation of the weight matrix yields $f(V_{W}^{(1)}(t)) \approx \chi_1 \chi_2 E_{W}^{(1)}(t)$.  Therefore from (main text) Equation 1, the resulting photoelectric multiplication is:

\begin{linenomath}
\begin{align}
V^{(1)}_{\text{out}}(t) &= \chi_{PD} \text{Im} \left[\left(\chi_1 \chi_2 E^{(j)}_X(t)\right)^* \chi_1 \chi_2 E^{(j)}_W(t)\right] \nonumber \\
&= \chi_{PD} \left( \chi_1 \chi_2 \right)^2 \text{Im} \left[\left(E^{(j)}_X(t)\right)^* E^{(j)}_W(t)\right] \nonumber,
\end{align}
\end{linenomath}

\noindent where $\chi_{PD}$ is determined by the responsivity of the photodetector and the termination resistance.

Hence for the linear characterization, we use a 1-parameter curve fit where the parameter estimates the value of $\chi_{PD} \left( \chi_1 \chi_2 \right)^2$.  To attain the curve fit parameter, we used a single randomized matrix-vector product and gradually increased the amplitude to create a curve, where the slope of the curve is determined by $\chi_{PD} \left( \chi_1 \chi_2 \right)^2$.  We re-calibrated the curve fit whenever we changed the size of the matrix-vector product being experimentally computed.  See Supplementary Section B for more details on the statistical linear curve fitting methods.

The (main text) Figure 3(d) shows the experimental matrix-vector multiplication performance of our architecture, where $Y$ is the expected curved-fitted value of the output vector, and $\hat{Y}$ is the experimental output vector.  Both $Y$ and $\hat{Y}$ are normalized to the largest value among all the products.  First, we characterized scalar-scalar products by computing 10,000 randomized scalar-scalar multiplications and comparing them to the curve fitted analytical product, yielding 9-bit precision.  Next, we computed 10,000 randomized $10\times10$ matrix-vector products to yield 8-bit precision.  Thus, we achieved accurate experimental linear matrix-vector products using this architecture.

\noindent \textbf{Experimental Methods:}  All experimental measurements of the neuron frequency modes throughout both the main text and supplementary used a spectrum analyzer (SA), other than the plots in the main text in Figure 2 and Figure 3(c), which used an oscilloscope.  Since SAs can only discern the magnitude of each frequency mode and not the phase, all measurements took the absolute value of the actual neuron value.  This does not prevent the experiment from computing negative-valued matrix algebra.

All the linear measurements with randomized values that determined a bit precision (Supplementary Figure \ref{fig:awg}(b) and main text Figure 3(d)) used a single readout from the SA (no averaging).  Additionally, the linear curve fits in Figures \ref{fig:lincurve}(b) and \ref{fig:lincurve}(c) also used a single readout per experimental data point.  The frequency correction measurements in \ref{fig:lincurve}(d) used the averaging function of the SA to average 100 samples, as that was a one-time deterministic characterization.

All curve fits were computed using the function ``curve\_fit'' from ``scipy.optimize'' in Scipy version 1.6.2, which provided both the curve fit parameters and the standard deviation errors.  Additionally, all curve fits are ``double-sided,'' allowing for both positive and negative deviations.

\subsubsection*{AWG Measurements}

\begin{figure}[h]
\includegraphics[width=0.75\linewidth]{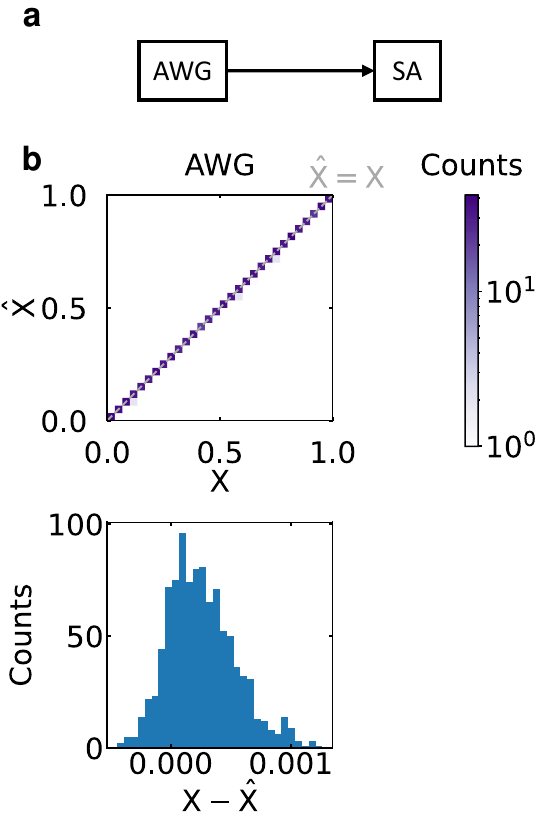}
\caption{Characterizing the bit-precision of the AWG.  (a)  The output of the AWG is directly measured by an SA.  (b)  A comparison between the expected value $X$ and the measured value $\hat{X}$ for 1,000 random scalars, where each scalar is simply the amplitude of a single frequency mode.  Above is a 2D histogram comparing the values of $X$ and $\hat{X}$, and below is a 1D histogram of the error $X - \hat{X}$.  This measurement yielded 12-bit precision for the AWG.} \label{fig:awg}
\end{figure}

For the linear curve fit, we set a baseline by characterizing the bit-precision of the AWG.  Figure \ref{fig:awg}(a) shows the experimental setup, where the AWG is directly connected to the SA.  This experiment consisted of generating single tone 20 MHz sine waves with known amplitudes $X$, measuring the amplitudes on the SA with a marker set on 20 MHz to yield $\hat{X}$, then normalizing $X$ and $\hat{X}$ to the largest value, and finally comparing $X$ with $\hat{X}$.  Note that SAs excel at relative frequency measurements, but require calibration for absolute measurements.  Thus, we first ran the measurement with 100 randomized values, then deterministically found the calibration factor by adjusting the linear scaling until the bit precision was maximized.  The SA calibration factor was found to be $\sim 0.68$.  We then ran the measurement again with this factor for 1,000 randomized values, where the results are shown in Figure \ref{fig:awg}(b).  We measured a normalized standard deviation of $2.8\cdot 10^{-4}$ to yield 12-bit precision.

\subsubsection*{Scalar-Scalar Product Measurements}

\begin{figure*}[t]
\centering
\includegraphics[width=\textwidth]{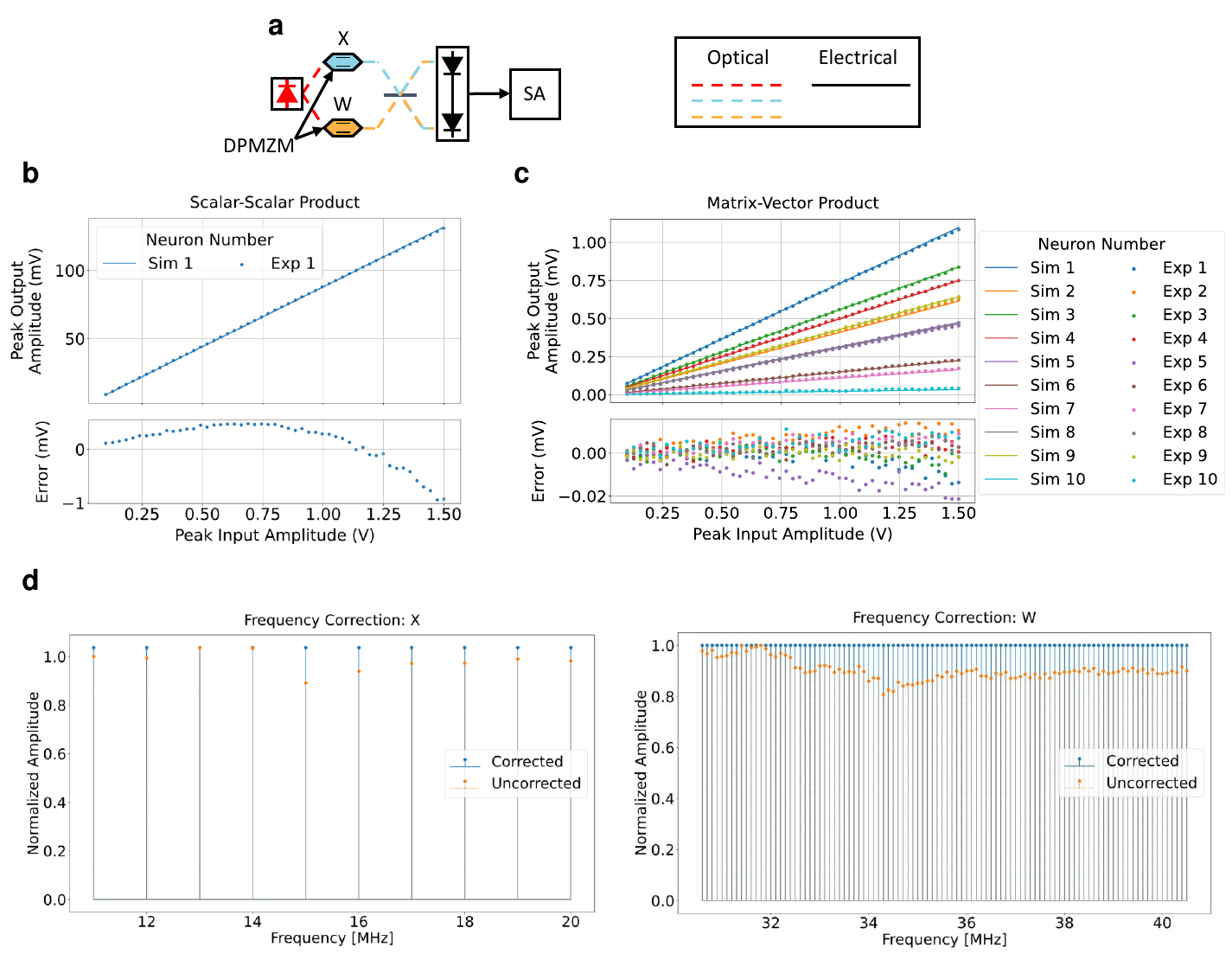}
\caption{The linear matrix-vector characterizations.  (a) The experimental setup for the linear characterization.  (b) The curve fit used for the scalar-scalar characterization in the main text.  (c) The curve fit used for the matrix-vector characterization in the main text.  (d)  The deterministic corrections to account for the uneven frequency response of the DPMZMs for the matrix-vector characterization in (c).}
\label{fig:lincurve}
\end{figure*}

Figure \ref{fig:lincurve}(a) shows the experimental setup used for the scalar-scalar and matrix-vector characterizations in the main text.  As described in the main text, we estimated the value of $\chi_{PD} \left( \chi_1 \chi_2 \right)^2$ using a 1-parameter curve fit for the linear characterizations.  We absorb the aforementioned SA calibration factor into all the curve fits.

The scalar-scalar product can be interpreted as a $1 \times 1$ matrix-vector multiplication, and thus is set up in the same way as the matrix-vector characterization. We used a single randomized scalar-scalar product to determine the curve fit.  The input vector $X$ was randomly set to a 20 MHz sinusoidal wave with peak amplitude of 0.8140314 V.  The weight matrix $W$ was randomly set to -0.88271151 V on a 2 MHz sinusoidal wave.  Thus the output $Y$ was a 18 MHz sinusoidal wave.  Since the AWG has a setting to scale the amplitudes of the signals, we gradually increased the amplitude of $X$ to create the curve in Figure \ref{fig:lincurve}(b).  The standard deviation error of this curve fit is $8.87\cdot10^{-5}$, where the units of the output signal is in volts.

Thus, we accurately computed 10,000 randomized products using a single curve fit from a random scalar-scalar product.  This measurement yielded a normalized standard deviation of $2.01\cdot 10^{-3}$, corresponding to 9-bit precision.

\subsubsection*{Matrix-Vector Product Measurements}

The experimental setup in Figure \ref{fig:lincurve}(a) was also used for the matrix-vector product measurements.  In this case, the input vector $X$ was a 10-frequency signal from 11 MHz to 20 MHz spaced at 1 MHz.  The weight signal $W$ was a 100-frequency signal from 30.6 MHz to 40.5 MHz spaced at 100 kHz.

For this measurement we included a deterministic correction for the uneven frequency response of the system.  After testing different photodectors and AWGs, we found that it was the DPMZMs that have a significantly varying frequency response even within kHz bandwidth.  We individually characterized each DPMZM in the setup illustrated in Figure \ref{fig:lincurve}(a) by using the AWG to send a flat RF frequency comb at the frequencies of interest to the DPMZM under test, and then sending a 100 MHz single tone signal to the other DPMZM.  This yields an output where the RF frequency comb is simply shifted by 100 MHz, but the shape of the output RF frequency comb reveals the frequency response of the DPMZM under test.

Figure \ref{fig:lincurve}(d) shows the result of these measurements.  The orange stems are the experimentally measured relative amplitudes of the frequency modes, characterizing the frequency response of the DPMZM.  The blue stems are all set to a normalized amplitude of exactly 1 to contrast the orange stems.  To apply this correction to the experiment, we multiply $X$ and $W$ by these frequency corrections before generating their signals from the AWG.

With these frequency corrections, we then used a randomized $10\times 10$ matrix-vector product for the curve fit in Figure \ref{fig:lincurve}(c).  This 1-parameter curve fit yielded a standard deviation error of $2.71\cdot 10^{-7}$, where the output signal was in units of volts.  At this point there is one last deterministic calibration required, which is to account for the scaling of the AWG.  That is, we must scale the peak amplitude of every signal to 1 V before sending it through the AWG (which then has its own amplitude setting).  Although this scaling factor is different for every signal, it is deterministic and known beforehand.  Thus, we multiply the curve fit parameter found from Figure \ref{fig:lincurve}(c) by this scaling factor to find the true parameter of the system.

Then, as explained in the main text, we compute 10,000 randomized $10\times 10$ matrix-vector products, where we apply the AWG scaling to each matrix-vector product.  This leads to a normalized standard deviation of $6.43\cdot 10^{-3}$, corresponding to the 8-bit precision presented in the main text.

\subsection{Nonlinear Curve Fit Characterizations}

\begin{figure*}[t]
\centering
\includegraphics[width=\textwidth]{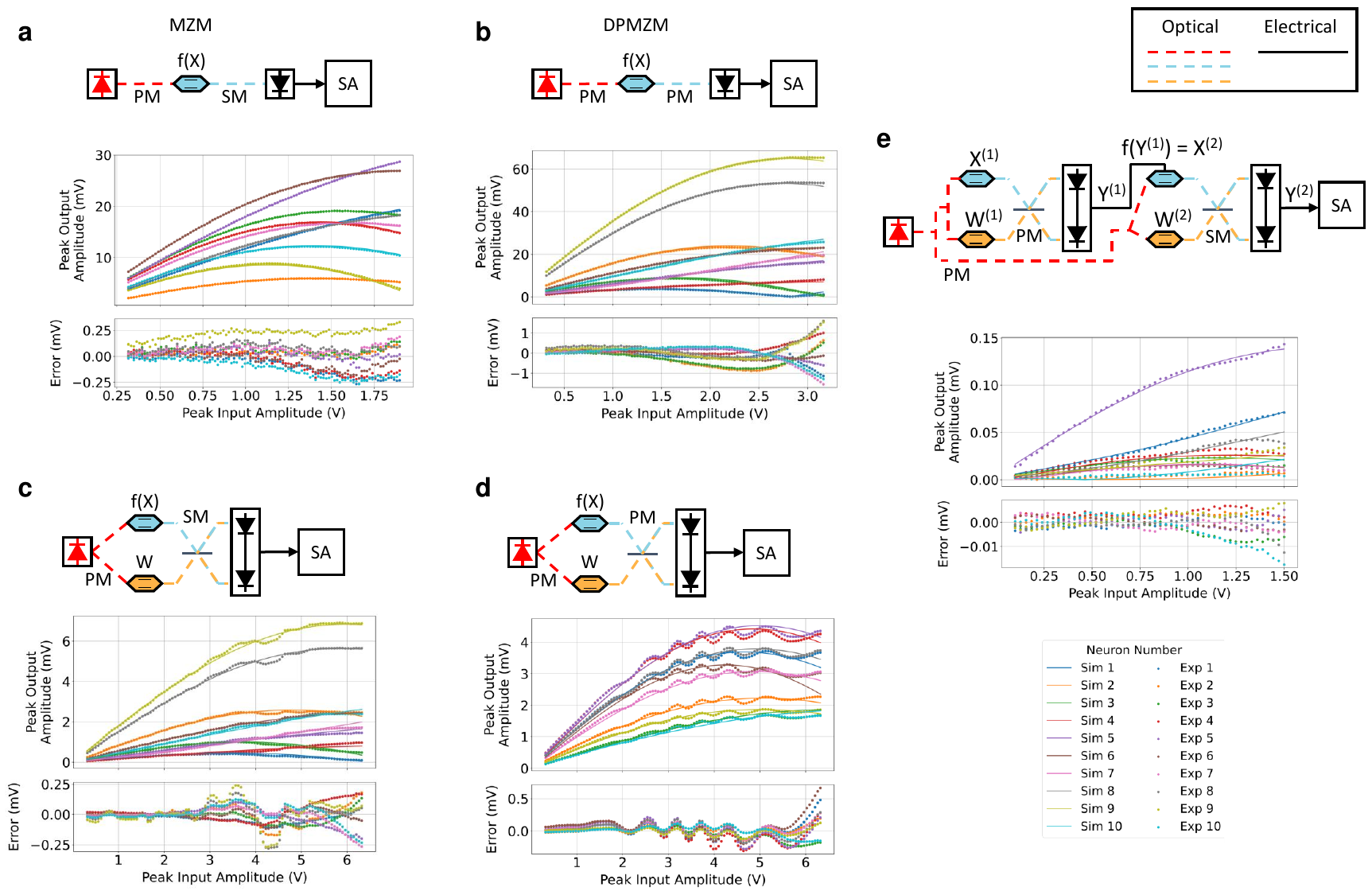}
\caption{Various configurations for characterizing the MZM nonlinearity.}
\label{fig:nonlin}
\end{figure*}

\begin{figure*}[t]
\centering
\includegraphics[width=\textwidth]{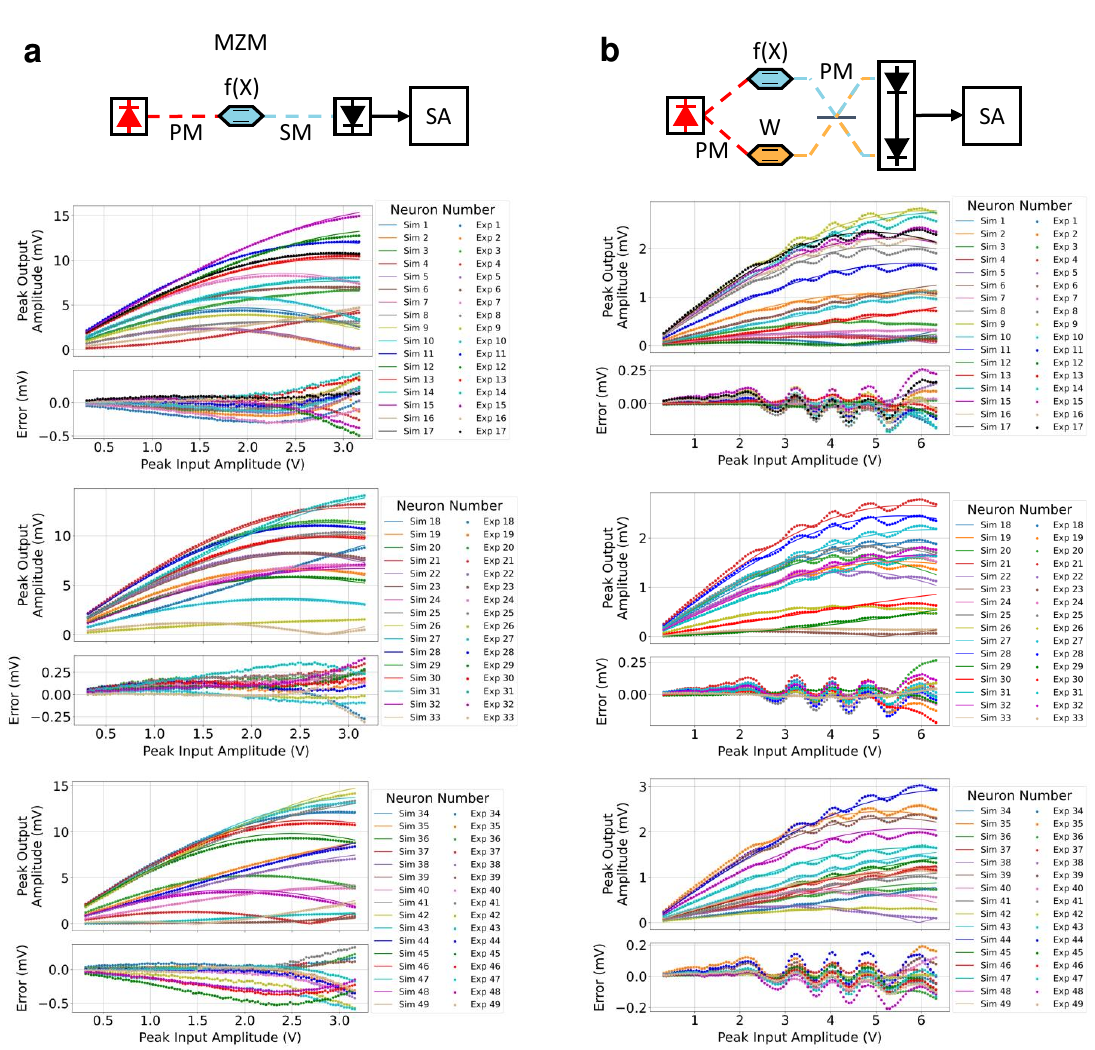}
\caption{A 49-frequency (size of an downsampled $7 \times 7$ MNIST image) nonlinear characterization, where we plot the nonlinear curves of all 49 neurons for two configurations from Figure \ref{fig:nonlin}.}
\label{fig:bignonlin}
\end{figure*}

\begin{figure*}[t]
\centering
\includegraphics[width=\textwidth]{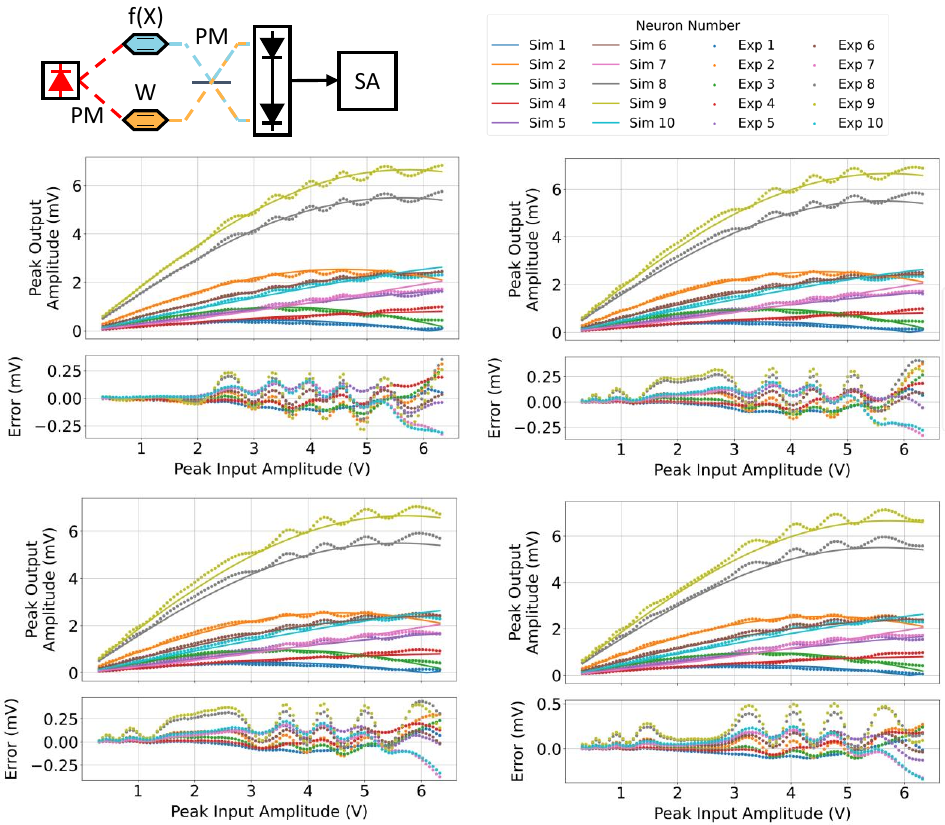}
\caption{A nonlinear measurement that was repeated four times to explore the nature of the ripples observed in Figure \ref{fig:nonlin}(d).  The curve fit parameters are the same for all four experiments and were derived from the top left measurement.}
\label{fig:squiggle}
\end{figure*}

Note that the following mathematical analysis uses the notation from Figure 3 in the main text.  All references to equations and figures to the main text will be explicitly stated.  All data contains real-valued positive and negative numbers.

\noindent \textbf{Mathematical Analysis:}  The (main text) Figure 3(e) illustrates a nonlinear curve fit for a simple intensity-modulated direct detection (IMDD) link, which consists of an electrical input signal $V^{(1)}_X(t)$ being modulated by an MZM and then immediately detected with a photodetector.  Our equation that models the output of the IMDD link is:

\begin{linenomath}
\begin{align*}
V_{\text{out}}(t) = \chi_0 + \chi_{PD} \chi_1 \sin \left(\chi_2 V_{X}^{(1)}(t) + \chi_3 \right).
\end{align*}
\end{linenomath}

Thus, we use a 4-parameter curve fit for the nonlinear characterization (counting $\chi_{PD} \chi_1$ as a single parameter).  The (main text) Figure 3(e) shows an example of curve fitting the analytical model to an experimental characterization of an MZM, where $V^{(1)}_X(t)$ is a $10\times1$ input vector.

\noindent \textbf{Experimental Methods:}  We also characterized the nonlinear behavior of modulators from various manufacturers and configurations to show the accuracy and applicability of our physics-based DNN model.  All our nonlinearity characterizations are 4-parameter curve fits to the equation: 

\begin{linenomath*}
\begin{align*}
f(V_X(t)) = \chi_0 + \chi_1 \sin \left(\chi_2 V_X(t) + \chi_3 \right), \nonumber
\end{align*}
\end{linenomath*}

\noindent where $V_X(t)$ is the input vector signal.  Note that we absorb all physical effects into the four parameters above, including $\chi_{PD}$, the SA calibration factor, etc.  And like the linear characterizations, all the nonlinear curve fits are also ``double-sided.''

Figure \ref{fig:nonlin} shows several configurations for which we curve fitted the nonlinearity.  These measurements were performed using the same method as the linear curve fit; by using the AWG amplitude setting to gradually increase the amplitude of the input vector $X$ until it reached the nonlinear regime of the modulator under test.  Each configuration within Figure \ref{fig:nonlin} has a different randomized input vector.  The DNN algorithm in Supplementary Section E was adjusted to match each curve fit configuration.  All configurations except the one in Figure \ref{fig:nonlin}(e) used the frequency correction method described in Supplementary Section B.  An RF amplifier was used for all configurations to increase the power of the electrical signal to reach the nonlinear regime of the modulator under test.  Table \ref{table:err} shows the standard deviation error of the various nonlinear curve fits.  All measurements in Figures \ref{fig:nonlin}(a)-(d) used the averaging function of the SA to average 50 samples for each experimental data point.  The data points in the 3-layer DNN characterization measurement in Figure \ref{fig:nonlin}(e) did not use the SA averaging feature, instead using a single SA readout per experimental data point.

\begin{table}
\renewcommand\tabularxcolumn[1]{m{#1}}
\begin{tabularx}{0.9\linewidth} { 
  | >{\centering\arraybackslash}X 
  || >{\centering\arraybackslash}X 
  | >{\centering\arraybackslash}X 
  | >{\centering\arraybackslash}X
  | >{\centering\arraybackslash}X |}
    \hline
    \textbf{Figure} & $\chi_0$ Error & $\chi_1$ Error & $\chi_2$ Error & $\chi_3$ Error \\
    \hline
    4(a) &  $4.52 \cdot 10^{-4}$ & $1.05 \cdot 10^{-5}$ & $7.71 \cdot 10^{-4}$ & $1.91 \cdot 10^{-2}$ \\
    \hline
    4(b) &  $1.36 \cdot 10^{-3}$ & $4.67 \cdot 10^{-4}$ & $1.33 \cdot 10^{-3}$ & $3.37 \cdot 10^{-2}$\\
    \hline
    4(c) & $7.01 \cdot 10^{-5}$ & $1.41 \cdot 10^{-5}$ & $6.24 \cdot 10^{-4}$ & $8.65 \cdot 10^{-3}$\\
    \hline
    4(d) & $5.73 \cdot 10^{-5}$ & $1.66 \cdot 10^{-5}$ & $1.17 \cdot 10^{-3}$ & $9.34 \cdot 10^{-3}$\\
    \hline
    4(e) & $1.11 \cdot 10^{-2}$ & $4.00 \cdot 10^{-4}$ & $6.17 \cdot 10^{-2}$ & $2.08 \cdot 10^{-2}$\\
    \hline
    5(a) & $2.89 \cdot 10^{-4}$ & $2.48 \cdot 10^{-5}$ & $7.31 \cdot 10^{-4}$ & $1.43 \cdot 10^{-2}$\\
    \hline
    5(b) & $1.66 \cdot 10^{-4}$ & $6.25 \cdot 10^{-5}$ & $1.02 \cdot 10^{-4}$ & $2.05 \cdot 10^{-2}$\\
    \hline
    6 & $8.89 \cdot 10^{-5}$ & $1.69 \cdot 10^{-5}$ & $1.77 \cdot 10^{-3}$ & $1.13 \cdot 10^{-2}$\\
    \hline
\end{tabularx}
\caption{The one standard deviation errors of the nonlinear curve fits for the various configurations, where the units of the output signal is volts.}
\label{table:err}
\end{table}

Figure \ref{fig:nonlin}(a) is a simple IMDD link, using a regular MZM biased at quadrature (not a DPMZM).  All other configurations exclusively use DPMZMs.  This MZM has a polarization-maintaining (PM) fiber lead input and a single-mode (SM) fiber lead output.  Thus, all optical fibers before the MZM are PM, and the ones after are SM.  The input vector signal $V_X(t)$ contained 10 frequencies from 100 MHz to 109 MHz spaced at 1 MHz.  A single (non-balanced) photodetector was used here.

Figure \ref{fig:nonlin}(b) has the same configuration as that in Figure \ref{fig:nonlin}(a), except the MZM is replaced with a DPMZM with a PM output fiber lead.  For all configurations with DPMZMs, we characterize the nonlinearity by driving only one of the sub-MZMs.  In this case, we used the bias controller to set the DPMZM to the SSB-SC mode that sets both sub-MZMs to their minimum bias point.  Then we manually changed the bias point of the sub-MZM under test to quadrature.  The frequencies of $V_X(t)$ are the same as in \ref{fig:nonlin}(a), but with a different set of randomized values.  This is the nonlinear characterization shown in the main text.

Figure \ref{fig:nonlin}(c) uses an interferometer configuration with two DPMZMs, testing the nonlinear behavior with photoelectric multiplication.  Here, the DPMZMs have an SM fiber output lead, and thus all fibers after the DPMZMs are SM.  As explained in Supplementary Section A, all sub-MZMs were biased at their minimum point.  Here, $V_X(t)$ is a 10-frequency signal from 10 MHz to 19 MHz spaced at 1 MHz with randomized values.  And similar to the linear characterization, we set $V_W(t)$ to be a single tone at 50 MHz, measuring the output signal from 60 MHz to 69 MHz.

Figure \ref{fig:nonlin}(d) has the same hardware configuration and parameters as (c), except that we used DPMZMs with PM fiber output leads, and thus used PM fibers throughout the whole configuration.  We consistently observed a ripple with this configuration, which is discussed later with Figure \ref{fig:squiggle}.  This is same the configuration used for the linear characterization in Supplementary Section B (excluding the RF amplifier).

Figure \ref{fig:nonlin}(e) is the same configuration as the 3-layer experiment in the main text.  This figure shows the curve fit that was used to infer the 200 MNIST images.  The SA measurements for both the curve fit and the MNIST inference experiment in the main text were single readouts without averaging.  For this curve fit, the values of $X^{(1)}$, $W^{(1)}$, and $W^{(2)}$ were all randomized.  For the MNIST inference and DNN training, the nonlinearity was programmed at the strongest setting (the right-most set of data points).

Figure \ref{fig:bignonlin} repeats the measurement for two of the configurations from Figure \ref{fig:nonlin}, except setting $V_X(t)$ to be a 49-frequency signal (the same size as a downsampled $7\times 7$ MNIST image) from 10.1 MHz to 14.9 MHz with 100 kHz spacing, with randomized values.  The configuration in Figure \ref{fig:bignonlin}(b) again sets $V_W(t)$ to be a single tone at 50 MHz.  We plot the nonlinear curves of all 49 neurons, thus demonstrating a larger scale nonlinearity.

Figure \ref{fig:squiggle} explores the ripples observed in the nonlinear characterization of the hardware configuration in Figure \ref{fig:nonlin}(d).  Here, the curve fit was only applied to the top left measurement, and the same parameters were used for the other three measurements.  All four measurements were taken within 40 minutes.  Here, we see that the number and position of the ripples vary per measurement.  We hypothesize that the ripples are caused either by (i) interference due to imperfect attenuation of the laser carrier and sideband that allows for SSB-SC modulation or (ii) optical path length imbalance in the interferometer.  If the cause is (i), then this can perhaps be rectified by achieving SSB-SC modulation using a passive optical filter or a different bias control scheme.  For (ii), then introducing efforts to balance the optical paths may help.

\noindent \textbf{Modulation Classification:}  The curve-fit for the 3-layer modulation classification used a similar curve fit method but a slightly different setup.  Here five curve fit parameters were used:

\begin{linenomath}
\begin{align*}
V^{(1)}_{\text{out}}(t) = \chi_0 + \chi_1 \sin \left(\chi_2 V_{X}^{(1)}(t) + \chi_3 \right) \\
V^{(2)}_{\text{out}}(t) = \chi_{4} \text{Im} \left[\left(E^{(2)}_X(t)\right)^* E^{(2)}_W(t)\right].
\end{align*}
\end{linenomath}

Using the new parameter $\chi_4$ does not introduce any new free variables in the curve fitting model but does enable the curve fitting algorithm to potentially find a more realistic fit.  The curve fitting parameters used in the modulation classification experiment in (main text) Figure 4(c) are:  $\chi_0=2.832$ with 1-STD error 550,320; $\chi_1=1.097$ with 1-STD error 213,189; $\chi_2=-3.409$ with 1-STD error 0.161; $\chi_3=0.236$ with 1-STD error 0.049; and $\chi_4=0.409$ with 1-STD error 79,569.

In addition instead of curve-fitting using the increasing-amplitude versus frequency mode magnitude curve as in the other nonlinear figures, we curve fitted the parameters using the FFT of $V^{(2)}_{\text{out}}(t)$ itself.  Although the curve fit parameters were accurate enough to achieve 85.0\% experimental accuracy for modulation classification, the high errors suggest that there may be a device imperfection or a gap in the physics model of the hardware.

\subsection{Spectrum Control with Matrix Operations}

\begin{figure*}[!ht]
\centering
\includegraphics[width=0.9\linewidth]{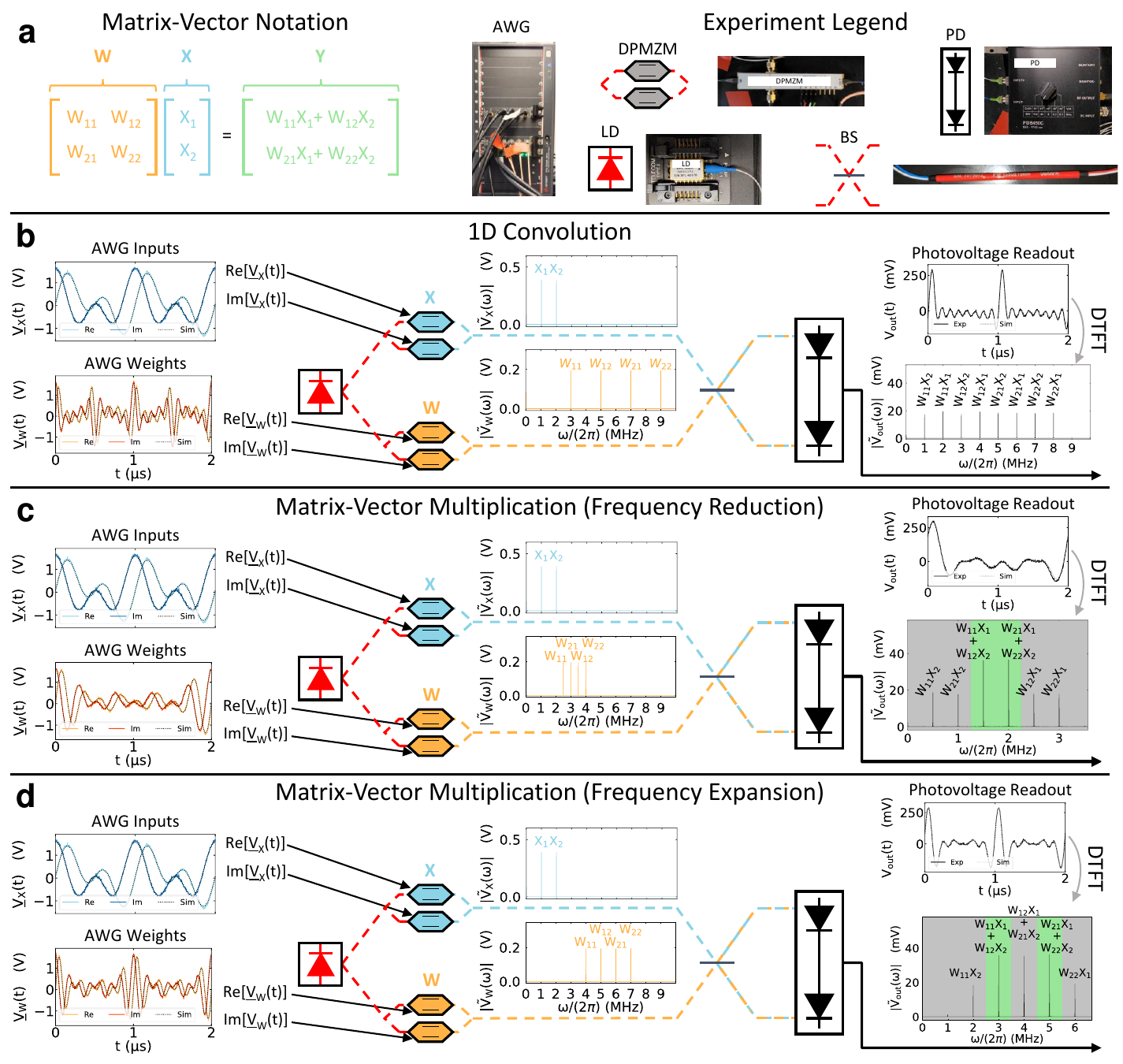}
\caption{An experimental example of various methods of computing using the MAFT scheme.  Each time domain plot shows one period of the raw data from the oscilloscope, and each frequency domain plot shows the discrete time Fourier transform (DTFT) of the entirety of the oscilloscope trace.  (a)  The notation for a typical $2\times2$ matrix multiplication between a matrix $W$ and a vector $X$.  The legend contains photos of the commercial components used to conduct the experiments in this figure.  (b) 1D Convolution.  In this case, the frequencies of $V_W(t)$ were chosen to demonstrate a separation of the eight partial sums contained in the frequencies of $V_{\text{out}}(t)$.  (c)  Frequency reduction scheme.  $V_W(t)$ is programmed to yield a matrix-vector product where $\Delta \omega_Y < \Delta \omega_X$.  Here, the spurious frequencies (gray region) are pushed to either side of the neuron frequencies.  The unique partial sum terms contained in the spurious frequencies of $V_{\text{out}}(t)$ will either be removed with a bandpass filter or used to train a DNN.  (d)  Frequency expansion scheme.  Here, $V_W(t)$ is programmed so that $\Delta \omega_Y > \Delta \omega_X$.  As with the previous scheme, the spurious frequencies in $V_{\text{out}}(t)$ will either be removed with a periodic filter or used to train the DNN.  The two frequency encoding schemes can be used alternatively to avoid bandwidth limitations for arbitrarily deep DNNs.}
\label{fig:biggie}
\end{figure*}

The $2 \times 2$ matrix analysis in the main text conveyed one scheme of programming a matrix-vector multiplication with MAFT.  However, MAFT also offers flexibility in determining the output spectrum of the matrix-vector products.  Figure \ref{fig:biggie} shows various methods of programming the spectrum of the output.

In this example, $V_X(t)$ contains two frequencies, $V_W(t)$ contains four frequencies, and the output electrical voltage signal $V_{\text{out}}(t)$ contains a variable number of frequencies.  We keep $V_X(t)$ the same while altering $V_W(t)$ to demonstrate various effects.  In Figure \ref{fig:biggie}(b), $V_W(t)$ is programmed so that each partial sum in $V_{\text{out}}(t)$ maps to a unique frequency, thus performing no summation in the frequency domain but instead performing a 1D convolution.  In Figure \ref{fig:biggie}(c), $V_W(t)$ is programmed so that the frequency domain of $V_{\text{out}}(t)$ yields a matrix-vector product where the frequencies corresponding to the elements of $Y$ are adjacently spaced in the middle of the spurious frequencies.  We refer to this method of programming as our \emph{frequency reduction scheme} because $\Delta \omega_Y < \Delta \omega_X$.  Figure \ref{fig:biggie}(d) demonstrates an alternative method of programming $V_W(t)$, which intersperses the elements of $Y$ with the spurious frequency components.  We term this the \emph{frequency expansion scheme}, as $\Delta \omega_Y > \Delta \omega_X$.  The frequency reduction and expansion schemes can be used alternatively for consecutive layers of a DNN to avoid running out of bandwidth.

Each partial sum term in $V_{\text{out}}(t)$ in Figures \ref{fig:biggie}(b)-(d) can be traced to difference between the input and weight frequencies.  For example, in Figure \ref{fig:biggie}(b) the partial sum term $W_{22}X_1$ derives from the product of $W_{22}$ at 9 MHz and $X_1$ at 1 MHz, where the difference between their frequencies causes $W_{22}X_1$ to appear at 8 MHz.

\subsection{Offline Physics-Based DNN Training}
For the 3-layer DNN MNIST inference experiment, the DNN was trained offline using an analytic model of the hardware.  As explained in the main text, four parameters were curve fitted to the experimental hardware.  One challenge was to create a DNN training algorithm that accurately models the photoelectric multiplication and nonlinearity while being fast enough to train the DNN in a reasonable time.

Pytorch was used for the DNN training, as the software automatically calculates the gradients for training the parameters, as long as only Pytorch functions are used to model the physics.  Algorithm 1 shows the pseudocode to quickly model the physics of the system.  We briefly explain each of the steps:

\begin{algorithm}
\caption{Physics-Based DNN Training}
\scriptsize
\begin{algorithmic}[1]
\State $\Delta f \gets \text{GCD}(\text{List of Expected Frequencies})$
\State $\widetilde{X}^{(1)}(f), \widetilde{W}^{(1)}(f), \widetilde{W}^{(2)}(f) \gets \text{DNN Matrix Values}$
\State $X^{(1)}(t), X^{(1)}_{\text{shift}}(t) \gets \text{DCT}(\widetilde{X}^{(1)}(f)), \text{DCT}_{\text{variant}}(\widetilde{X}^{(1)}(f))$
\State $W^{(1)}(t), W^{(1)}_{\text{shift}}(t) \gets \text{DCT}(\widetilde{W}^{(1)}(f)), \text{DCT}_{\text{variant}}(\widetilde{W}^{(1)}(f))$
\State $W^{(2)}_{\text{shift}}(t) \gets \text{DCT}_{\text{variant}}(\widetilde{W}^{(2)}(f))$
\State $Y^{(1)}(t) \gets X^{(1)}(t)W^{(1)}_{\text{shift}}(t) -  X^{(1)}_{\text{shift}}(t)W^{(1)}(t)$
\State $f(Y^{(1)}(t)) \gets \chi_0 + \chi_1 \sin \left(\chi_2 Y^{(1)}(t) + \chi_3 \right)$
\State $X^{(2)}(t) \gets f(Y^{(1)}(t))$
\State $Y^{(2)}(t) \gets X^{(2)}(t)W^{(2)}_{\text{shift}}(t)$
\State $\widetilde{Y}^{(2)}(f) \gets \text{DCT}(Y^{(2)}(t))$
\end{algorithmic}
\end{algorithm}
\normalsize

\begin{enumerate}
  \item Taking the greatest common denominator (GCD) of some frequencies we expect to be present to discretize the frequency space
  \item Directly inserting the vector and matrix values into the frequency domain
  \item Converting the inputs to the time domain, where the "shift" of the signal is the $90^{\circ}$ shifted signal used for SSB-SC modulation, and the "variant" of the discrete cosine transform (DCT) is an altered version of the DCT that yields the desired $90^{\circ}$ shifted signal
  \item Converting the layer 1 weights to the time domain
  \item Converting the layer 2 weights to the time domain
  \item The linear output of the first photoelectric multiplication; this signal contains the spurious frequencies, which can be filtered out at this point if desired
  \item Applying the curve fitted nonlinear activation
  \item The nonlinear output of the first layer is the input to the second layer
  \item The second photoelectric multiplication, modeling the photoelectric multiplication of the input in the DSB-SC mode and the weights in the SSB-SC mode
  \item Converting the output to the frequency domain to retrieve the neuron values
\end{enumerate}

We also included the effects of the AWG scaling and the frequency correction mentioned in Supplementary Section B into the algorithm, depending on the hardware configuration being modeled.  And although we did not use any frequency filters for our experiments, we incorporated the effects of attenuation from a filter into the DNN model.

The physics-based DNN training was found to take up to approximately 5-6 times longer to train compared to a conventional DNN, depending on the workload.  A better optimized physics-based algorithm should be able to cut the time down.

Note that this algorithm avoids using complex numbers to model the hardware or train the DNN.  However, the training for the modulation classification used a similar algorithm but one that included complex numbers so that both the magnitudes and phases of the weight signal could be trained.  Hence, the algorithm above is useful for scenarios where complex numbers are best avoided when training the DNN.

\subsection{LTI Experiments}

The three LTI signal processing experiments used the hardware setup shown in Figure \ref{fig:nonlin}(d) (though operating in the linear regime of the DPMZMs).  As explained in Supplementary Section A, each DPMZM has three DC biases.  To control whether the left or right sideband suppressed in SSB-SC operation, we programmed the polarity of the quadrature setting (positive versus negative quadrature) on the bias controller.   (The `left' sideband is the sideband with frequency lower than the laser carrier and the `right' sideband has higher frequency than the laser carrier.)

The LTI experiments require the phase information in the final readout of the FFT of the output waveform so that we can distinguish between the positive and negative frequencies.  Therefore instead of using a spectrum analyzer or taking the magnitude of the FFT from an oscilloscope, we took the imaginary part of the FFT to retain the phase information.  We inferred the correct timing of each waveform with a time domain matched filter and by observing the gradient and location of the phases (as a time delay deterministically introduces a phase gradient).

\begin{figure}[h]
\includegraphics[width=0.75\linewidth]{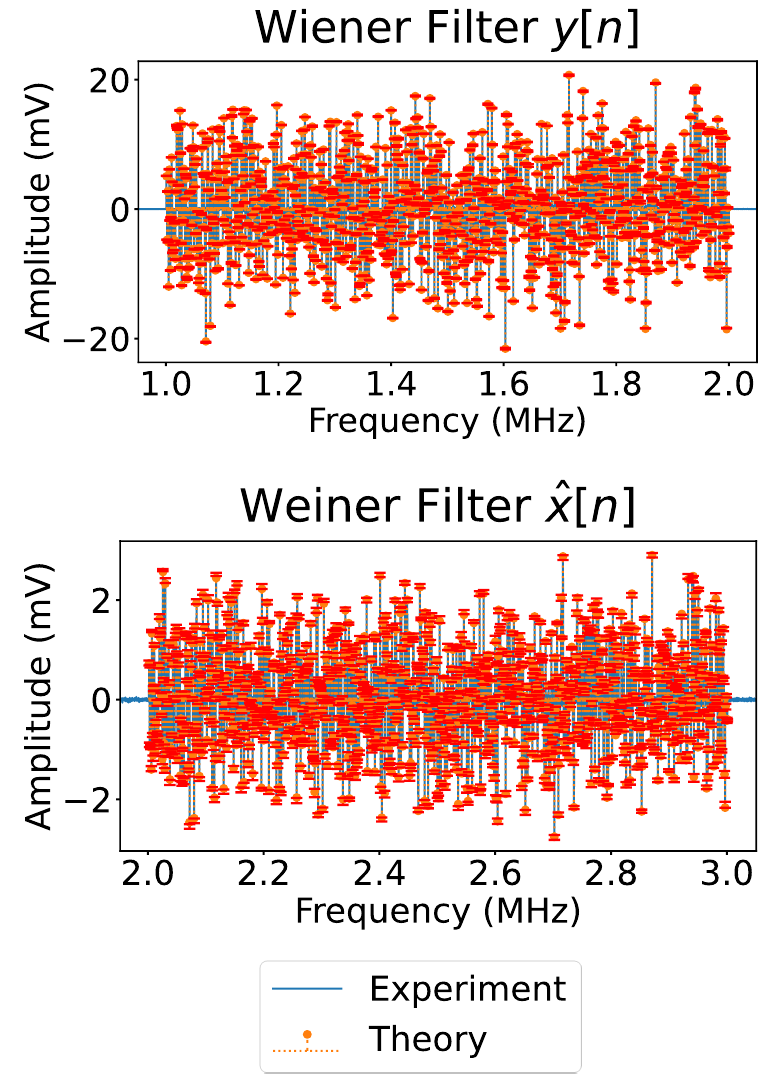}
\caption{Example experimental LTI measurements.  Each plot overlays the experiment and the theoretical signal in the Wiener filter scenario.  The top measurement compares the experimental generation of $y[n]$ from the AWG to the theoretical signal.  The bottom measurement compares the output of MAFT-ONN of $\hat{x}[n]$ to the theoretical signal.  The experimental waveform plotted is the mean of 10 experiments and the red error bars are the standard deviation of those experiments.} \label{fig:ltiexp}
\end{figure}

Each LTI experimental waveform was repeated 10 times on the MAFT-ONN hardware.  The MSEs were calculated by normalizing the experimental LTI waveforms with the theoretical LTI signals to account for the output scaling due to MAFT-ONN physics, ensuring that the signals were properly compared.  In (main text) Figures 4(a)-(c), the signals $x[n]$, $y[n]$, $h[n]$, and $x_{\text{target}}[n]$ were measured directly from the AWG while the signals $\hat{x}[n]$, $x_1[n]$, $x_2[n]$, and $x_3[n]$ were measured outputs of MAFT-ONN.  Figure \ref{fig:ltiexp} shows an example of an overlay of the theoretical versus experimental LTI signals where the experimental waveform is the mean of 10 experiments and the error bars are the standard deviation.

\section*{Theory}
\subsection{General MAFT Algorithm}
The input and output frequencies do not have to be evenly spaced like they are in the main text.  We used the even frequency spacing in the main text to maximize the throughput, which is limited by the time required to resolve the smallest frequency spacing.  This general frequency-encoding algorithm will show that a set of input frequencies can be transformed to an arbitrary set of output frequencies while computing an arbitrary matrix-vector product $W X = Y$, where $W$ has size $(R\times N)$, $X$ has size $(N\times 1)$, and $Y$ has size $(R\times 1)$.  This can all be accomplished in a single shot in the frequency domain.

As in the main text, we will assume that the input activation signal has frequency spacing $\Delta \omega_X$ and offset $n_0 \cdot \Delta \omega_X$; hence, the input vector begins as the electrical voltage signal $V_X(t)=\sum_{n=1}^{N}{X_{n} \cos((n_0 + n) \Delta \omega_X t)}$.  Here we assume equal frequency spacing for the input signal for convenience, without loss of generality.  The input signal can express an arbitrary set of frequencies by decreasing $\Delta \omega_X$ and setting the non-occupied frequencies to zero.  Now we do not assume anything about the output signal frequencies, leaving the weight frequencies as arbitrary values.  Therefore, let the frequency corresponding to the weight matrix element $W_{r,n}$ be $\omega_{r,n}^W$.  This yields the electrical voltage signal for the weight matrix: $V_W(t) = \sum_{r=1}^{R}\sum_{n=1}^{N}{W_{r,n} \cos(\omega_{r,n}^W t)}$.

Next, $V_X(t)$ and $V_W(t)$ are SSB-SC modulated onto a laser carrier with frequency $\omega_{\text{LD}}$.  This yields (ignoring linear scaling factors):

\begin{linenomath*}
\begin{align*}
 E_X(t) \propto \sum_{n=1}^{N}{X_{n} e^{i((n_0 + n) \Delta \omega_X + \omega_{LD})t}}, \\
 E_W(t) \propto  \sum_{r=1}^{R}\sum_{n=1}^{N}{W_{r,n} e^{i(\omega_{r,n}^W + \omega_{LD}) t}}.
\end{align*}
\end{linenomath*}

Next, the optical fields $E_X(t)$ and $E_W(t)$ enter a $2 \times 2$ 50:50 beam splitter and then a balanced photodetector for the photoelectric multiplication.  The optical outputs of the beam splitter are:

\begin{linenomath*}
\begin{align*}
E_{\text{BS1}}(t) & \propto E_X(t) - i E_W(t), \\
E_{\text{BS2}}(t) & \propto  -i E_X(t) + E_W(t).
\end{align*}
\end{linenomath*}

Then $E_{\text{BS1}}(t)$ and $E_{\text{BS2}}(t)$ each enter a photodetector, and the outputs are subtracted to yield the photoelectric multiplication output (again ignoring scaling factors):

\begin{linenomath*}
\begin{align}
&V_{\text{out}}(t) \propto \left |E_{\text{BS1}}(t) \right |^2 - \left |E_{\text{BS2}}(t) \right |^2  \nonumber \\
& \propto \text{Im} \left[E_X^*(t)E_W(t)\right]  \nonumber \\
& \propto \text{Im} \left[\sum_{r=1}^{R}\sum_{n'=1}^{N}\sum_{n=1}^{N}{W_{r,n'}X_n e^{i((\omega_{r,n'}^W - (n_0+n) \Delta \omega_X) t)}} \right]  \nonumber \\
& \propto \sum_{r=1}^{R}\sum_{n'=1}^{N}\sum_{n=1}^{N}{W_{r,n'}X_n \sin((\omega_{r,n'}^W - (n_0+n) \Delta \omega_X) t)} \label{eq:i}
\end{align}
\end{linenomath*}

Everything so far in this analysis has only considered the physics of the system.  Now we can choose the interpretation of the signals to represent matrix-vector product.  By definition of matrix multiplication, the output matrix elements are: $Y_r = \sum_{n=1}^{N}{W_{r,n}X_n}$.  By comparing this definition of a matrix product to Equation \ref{eq:i}, we see that only the frequencies that correspond to $n'=n$ will contribute to the desired operation.  All other frequencies will be spurious.  There are a total of $RN^2$ unique partial sums generated by this photoelectric multiplication.

This matrix-vector product can be achieved by programming the weight frequencies to group some of the $RN^2$  partial sums into $R$ groups of $N$ frequencies.  Each group is associated with a value $r$ and thus the element of the output matrix $Y_r$.  Say that we want to map each output matrix element $Y_r$ to frequency $\omega_r^Y$.  Then the solution for the weight matrix frequencies to achieve this desired frequency transformation is:

\begin{linenomath*}
\begin{align}
\omega_{r,n'}^W = \omega_r^Y + (n_0 + n') \Delta \omega_X, \label{eq:w}
\end{align}
\end{linenomath*}

 for $n' \in [1,\ldots,N], r \in [1,\ldots,R]$.  Plugging Equation \ref{eq:w} into Equation \ref{eq:i}, we have:
 
\begin{linenomath*}
\begin{align}
&V_{\text{out}}(t) \propto \nonumber \\
& \propto \sum_{r=1}^{R}\sum_{n'=1}^{N}\sum_{n=1}^{N}{W_{r,n'}X_n \sin((\omega_r^Y + (n'-n) \Delta \omega_X) t)} \label{eq:im} \\
& \propto \sum_{r=1}^{R}\sum_{n=1}^{N}{W_{r,n}X_n \sin(\omega_r^Y t)} \nonumber \\
& \ \ \ \ \ \ \ \ \ + \sum_{r=1}^{R}\sum_{n'=1}^{N}\sum_{\substack{n=1 \\ n \neq n'}}^{N}{W_{r,n'}X_n \sin((\omega_r^Y + (n'-n) \Delta \omega_X) t)} \nonumber \\
& \propto \sum_{r=1}^{R}{Y_r \sin(\omega_r^Y t)} \label{eq:terms} \\
& \ \ \ \ \ \ \ \ \ + \sum_{r=1}^{R}\sum_{n'=1}^{N}\sum_{\substack{n=1 \nonumber \\ n \neq n'}}^{N}{W_{r,n'}X_n \sin((\omega_r^Y + (n'-n) \Delta \omega_X) t)} \nonumber \\
& \propto V_Y(t) + V_S(t) \label{eq:sep},
\end{align}
\end{linenomath*}

\noindent where we ignore the linear scaling factor, which is discussed in the main text.  From Equation \ref{eq:terms} to  Equation \ref{eq:sep}, we group the terms with $n'=n$ as the output vector $Y$, described as $V_Y(t)$ and the terms with $n'\neq n$ as the `spurious frequency' components $V_S(t)$.

\textit{Thus, by programming the weight signal accordingly, we perform both a matrix-vector product and frequency transformation.}   The electric voltage output signal is $V_Y(t) = \sum_{r=1}^{R}{Y_r \sin(\omega_r^Y t)}$, and the remainder of the signal contains the spurious frequencies.  We arrive to the same equations in the main text by choosing equally spaced output vector frequencies:  $\omega_r^Y = (r_0 + r) \Delta \omega_Y$.

\subsection{Anti-Aliasing Conditions}
When computing matrix-vector products for a fully connected layer, the frequency reduction and expansion schemes from the main text use the result from Equation \ref{eq:terms} in a way that prevents aliasing.  This occurs when some of the spurious frequencies overlap with the output vector frequencies.  Each scheme avoids aliasing in a different way.  Recall that in both schemes, $\omega_r^Y = (r_0 + r) \Delta \omega_Y$, as in the main text.

\begin{figure*}[t]
\centering
\includegraphics[width=\textwidth]{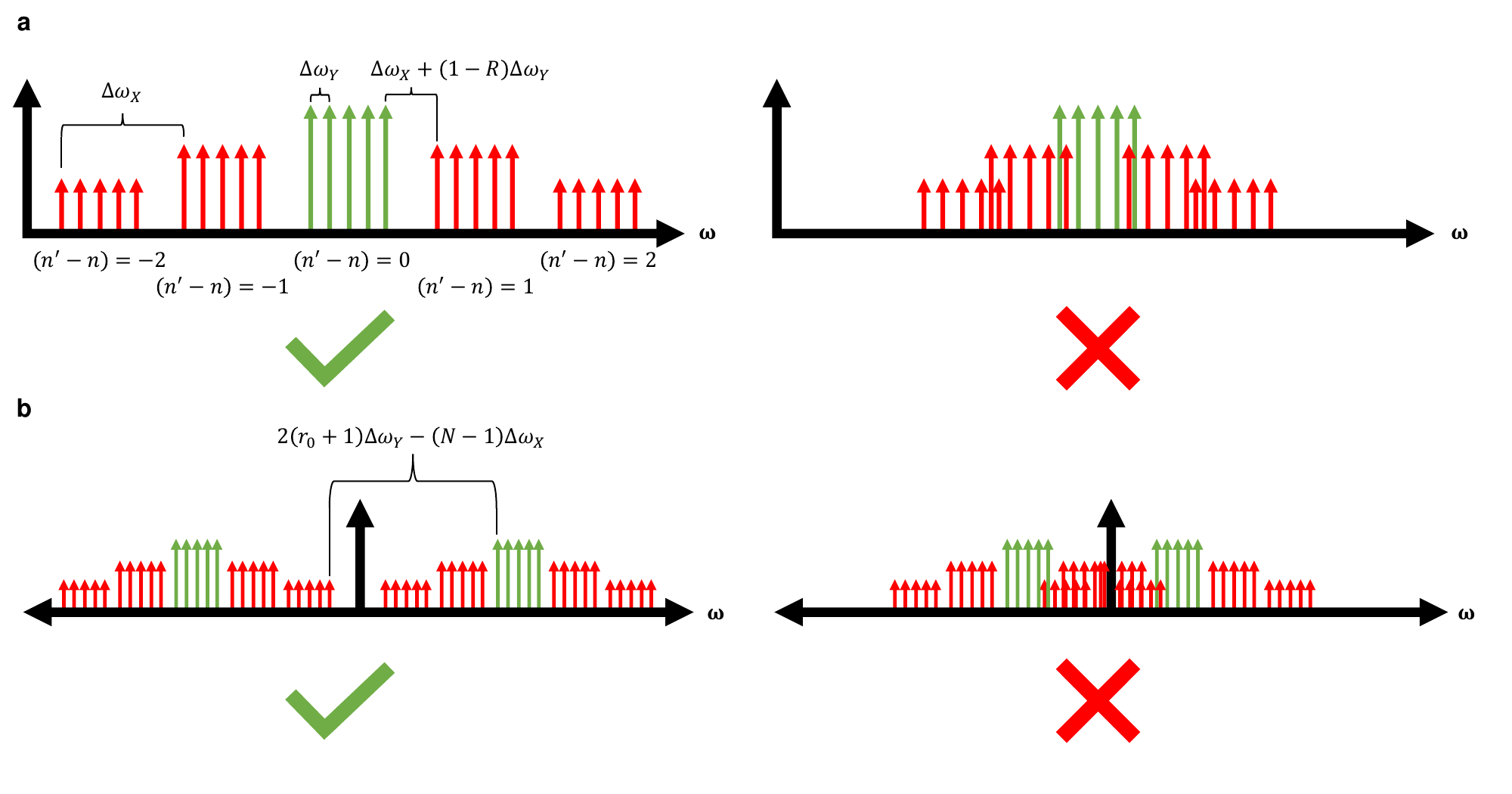}
\caption{The anti-aliasing conditions for the frequency reduction scheme, where the output vector frequencies are green and the spurious frequencies are red.  (a)  The first type of aliasing comes from the spurious frequencies overlapping with the output vector frequencies.  This can be avoided by decreasing the output frequency spacing.  (b)  The second type of aliasing comes from the negative frequencies creeping up to the positive region.  This can be avoided by increasing output frequency offset.}
\label{fig:red}
\end{figure*}

First, we analyze the anti-aliasing conditions of the frequency reduction scheme.  Here, the frequency spacing of the output frequencies are less than that of the input frequencies: $\Delta \omega_Y < \Delta \omega_X$.  As Figure \ref{fig:red} illustrates, there are two types of aliasing that must be avoided.

The first type of aliasing, shown in Figure \ref{fig:red}(a), originates from the overlapping of the spurious frequencies with the output vector frequencies.  We find the anti-aliasing constraint by keeping track of the distance of the closest spurious frequency to the edge of band of output vector frequencies.  From Equation \ref{eq:terms}, the lowest spurious frequency after the band is $(r_0+1)\Delta \omega_Y + \Delta \omega_X$, and the highest output vector frequency is $(r_0+R)\Delta \omega_Y$.  Thus, as labeled in Figure \ref{fig:red}(a) the gap between these frequencies is $\Delta \omega_X + (1-R)\Delta \omega_Y$.  Hence, for anti-aliasing, this gap must be greater than 0 at minimum.  This yields the constraint:  $\Delta \omega_Y < \frac{1}{R-1}\Delta \omega_X$.  A reasonable design choice, which is the one we made for the experiment, is to set this gap equal to the smallest frequency spacing already present in the signal, $\Delta \omega_Y$.  So with this anti-aliasing condition we get:  $\Delta \omega_Y = \frac{1}{R}\Delta \omega_X$.

The second anti-aliasing condition comes from the fact that the signal is real, and thus, in the frequency domain, it is possible for the negative part of the signal to creep into the positive region.  This phenomenon is conveyed in Figure \ref{fig:red}(b).  Here, we calculate the gap between the lowest output vector frequency and the least-negative spurious frequency.  The former is $(r_0+1)\Delta \omega_Y$.  The latter, according to Equation \ref{eq:terms}, is $-\left((1-N)\Delta \omega_X + (r_0 + 1)\Delta \omega_Y \right)$.  Therefore, as labeled in Figure \ref{fig:red}(b) the frequency gap is $2(r_0+1)\Delta \omega_Y - (N-1) \Delta \omega_X$.  Again, we want this gap to be greater than 0 to avoid aliasing, so we end up with the constraint $(r_0+1)\Delta \omega_Y > \frac{1}{2}(N-1)\Delta\omega_X$.  Another reasonable design choice is to also set this gap equal to $\Delta \omega_Y$.  Applying both design choices, we solve for the term $r_0$ to get:  $r_0\Delta \omega_Y = \frac{1}{2}(N-\frac{R+1}{R})\Delta\omega_X$.  Now we have a set of constraints and a way to program the weight signal frequencies to guarantee that the spurious frequencies do not overlap with the output vector frequencies when using the frequency reduction scheme.

\begin{figure*}[t]
\centering
\includegraphics[width=\textwidth]{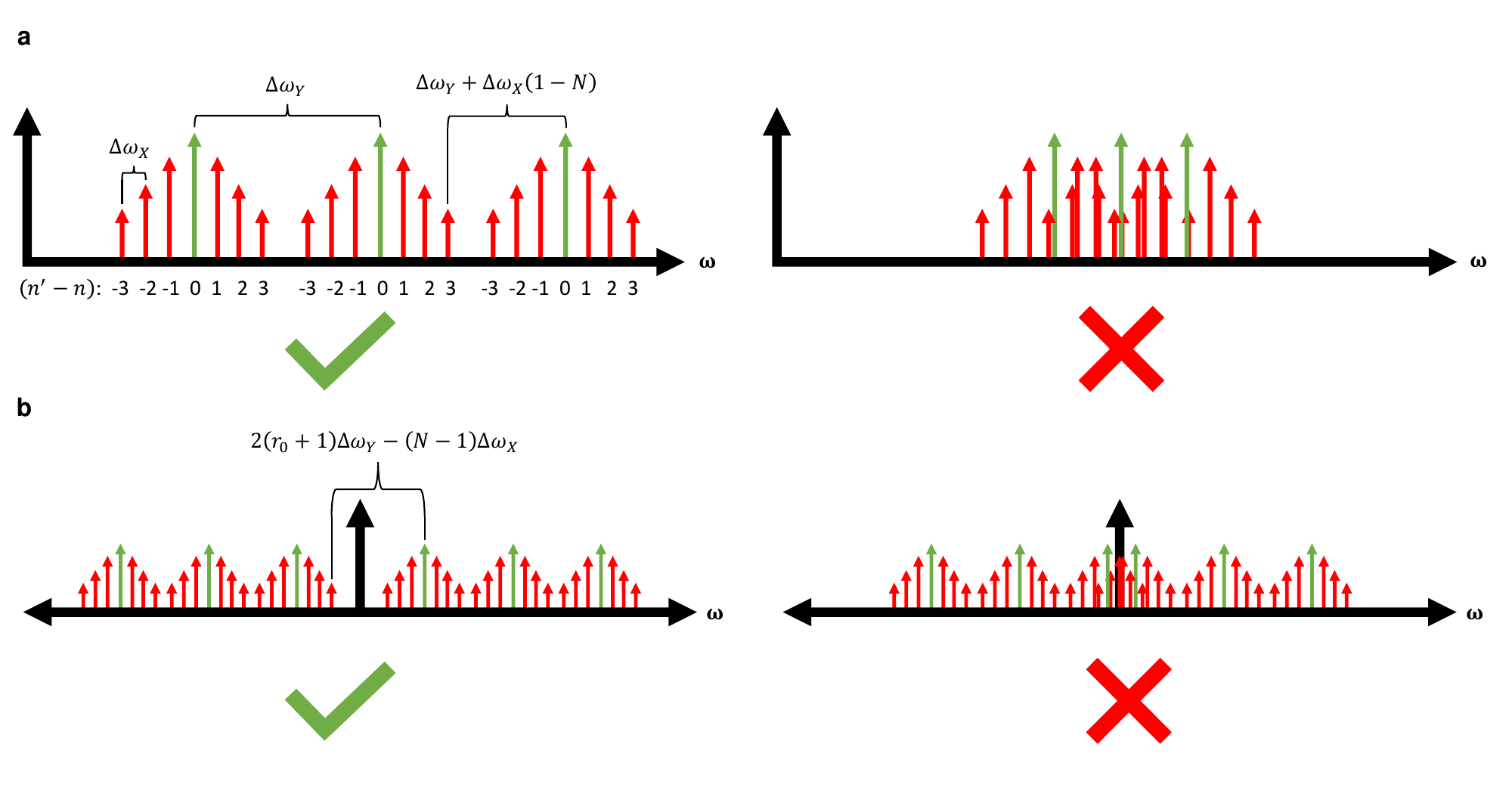}
\caption{The anti-aliasing conditions for the frequency expansion scheme, where the output vector frequencies are green and the spurious frequencies are red.  (a)  The first type of aliasing comes from the spurious frequencies overlapping with the output vector frequencies.  This can be avoided by increasing the output frequency spacing.  (b)  The second type of aliasing comes from the negative frequencies creeping up to the positive region.  This can be avoided by increasing output frequency offset.}
\label{fig:expa}
\end{figure*}

Next is the frequency expansion scheme, where the output vector frequency spacing is larger than the input frequency spacing; $\Delta \omega_Y > \Delta \omega_X$.  Figure \ref{fig:expa} illustrates the anti-aliasing conditions.  The two types of aliasing here are the same as in the frequency reduction scheme.

For the first type of aliasing, showed in Figure \ref{fig:expa}(a), we calculate the frequency gap between the $j^{\text{th}}$ output vector frequency and the closest spurious frequency.  The frequency of the $j^{\text{th}}$ output vector is $(r_0+j)\Delta \omega_Y$, and the closest spurious frequency is $(r_0+j-1)\Delta \omega_Y + (N-1) \Delta \omega_X$.  (The spurious frequencies are symmetric about each output vector frequency, so this calculation can also be done with the higher spurious frequency.)  Thus, the frequency gap is $\Delta \omega_Y - \Delta \omega_X (N-1)$.  Therefore, the anti-aliasing condition is:  $\Delta \omega_Y > \Delta \omega_X (N-1)$.  And another reasonable design choice is to set this gap equal to the smallest frequency spacing present in the signal, $\Delta \omega_X$.  With this, we get:  $\Delta \omega_Y = N \Delta \omega_X$.

Figure \ref{fig:expa}(b) shows the second type of aliasing for the frequency expansion scheme, where the negative frequencies of the real signal creep into the position region.  The constraint arising from this aliasing is derived in the same way.  The lowest output vector frequency is $(r_0+1)\Delta \omega_Y$, and the least-negative spurious frequency is $-\left((1-N)\Delta \omega_X + (r_0+1)\Delta \omega_Y \right)$.  Thus, the frequency gap is $2(r_0+1)\Delta \omega_Y - (N-1)\Delta \omega_X$, yielding the anti-aliasing constraint: $(r_0 +1) \Delta \omega_Y > \frac{1}{2}(N-1)\Delta \omega_X$.  Interestingly, when using the design choice from above, this constraint becomes $r_0 \Delta \omega_Y > -\frac{1}{2}(N-1)\Delta \omega_X$.  Since the constraint is a negative number, we can simply set the offset frequency for the output vectors to zero as a convenient design choice, so $r_0 = 0$.

As shown above, the spurious frequencies generated by the photoelectric multiplication requires anti-aliasing conditions to ensure they do not degrade the matrix-vector product.  In addition to the constraints, we also presented some convenient design decisions for handling the spectrum.  These design decisions do not have to be followed, however.  For example, for our experiment, in the layer 1 frequency reduction scheme, we chose a value of $r_0$ that was larger than necessary in order to push all the frequencies beyond 2 MHz, which is the lower limit of our RF amplifier.

\subsection{Throughput Derivation}

The throughput is a measure of the volume of data that can be processed within a given time.  Since this architecture directly computes matrix-vector products, we analyze the throughput of a single fully-connected layer of a DNN.  From the main text, the general throughput equation is:

\begin{linenomath*}
\begin{align}
T = \frac{\text{\# \ operations}}{\text{latency}} = N \cdot R \cdot \min(\Delta f, f_0),
\label{eq:tp}
\end{align}
\end{linenomath*}

\noindent where $N$ is the number of input neurons into the layer, $R$ is the number of output neurons, $\Delta f$ is the smallest spacing of the output signal, and $f_0$ is the lowest neuron frequency of the output signal.  Here we derive the throughput for the frequency reduction and expansion schemes, given the anti-aliasing conditions found in the previous section.

The available bandwidth with which to modulate the inputs and weights, $B$, limits the weight matrix signal, as that signal will always have a higher frequency than the input signal, as shown in Equation \ref{eq:w}.  Therefore, the maximum bandwidth used to modulate the signals is:  $B = (n_0 + N) \Delta f_X + (r_0 + R) \Delta f_Y$.  As we saw in Supplementary Section G, the term $n_0$ cancels out and does not affect the output frequencies.  So as a design choice to maximize the available bandwidth, we set $n_0=0$ to yield $B = N \Delta f_X + (r_0 + R) \Delta f_Y$.

For the frequency reduction scheme, it is clear that the limiting factor in the term $\min(\Delta f, f_0)$ is the frequency spacing of the output signal, $\Delta f_Y$.  This is because by definition the frequency reduction scheme requires that $\Delta f_Y < \Delta f_X$, and we see from the anti-aliasing conditions that $\Delta f_Y < r_0 \Delta f_Y$.  Hence, the throughput of the frequency reduction scheme is: $T_{\text{reduction}} = N \cdot R \cdot \Delta f_Y$.

Now we can solve for the throughput in terms of the bandwidth, using the design choices mentioned in the anti-aliasing section, as those choices maximize the throughput.  First, we can plug in the value of $\Delta f_Y$ into the throughput equation using the anti-aliasing result $\Delta f_Y = \frac{1}{R}\Delta f_X$ to yield:  $T_{\text{reduction}} = N \cdot \Delta f_X$.

Dividing the throughput by the bandwidth, we get:

\begin{linenomath*}
\begin{align*}
\frac{T_\text{reduction}}{B} = \frac{N \Delta f_X}{ N \Delta f_X + (r_0 + R) \Delta f_Y}.
\end{align*}
\end{linenomath*}

Here, we can again plug in the anti-aliasing values of $\Delta f_Y = \frac{1}{R}\Delta f_X$ and $r_0 \Delta f_Y = \frac{1}{2}(N-\frac{R+1}{R})\Delta f_X$, to yield:

\begin{linenomath*}
\begin{align*}
& \frac{N \Delta f_X}{ N \Delta f_X + (r_0 + R) \Delta f_Y} = \\
&= \frac{N \Delta f_X}{ N \Delta f_X + \frac{1}{2}\left(N-\frac{R+1}{R}\right) \Delta f_X  + \frac{\Delta f_X}{\Delta f_Y} \Delta f_Y} \\
&= \frac{N \Delta f_X}{ N \Delta f_X + \frac{1}{2}\left(N-\frac{R+1}{R}\right) \Delta f_X  + \Delta f_X} \\
&= \frac{N}{ N + \frac{1}{2}\left(N-\frac{R+1}{R}\right) + 1} \\
&= \frac{2NR}{3NR + R + 1},
\end{align*}
\end{linenomath*}

\noindent where the final equality came from simplifying the expression.  To get the final approximation for the throughput, we can take the reciprocal of this expression to see that:

\begin{linenomath*}
\begin{align*}
\frac{B}{T_\text{reduction}} &= \frac{3NR + R + 1}{2NR} \\
&= \frac{3}{2} + \frac{1}{2N} - \frac{1}{2NR} \\
& \approx \frac{3}{2}, \ \ \ \ \  \text{for} \ \ N \gg 1. \\
\end{align*}
\end{linenomath*}

Finally, taking the reciprocal again, we have both the exact and approximate expressions for the throughput of the frequency reduction scheme under the anti-aliasing conditions:

\begin{linenomath*}
\begin{align}
T_\text{reduction} &= \frac{2NR}{3NR + R + 1} B \approx \frac{2}{3} B
\label{eq:tpr}.
\end{align}
\end{linenomath*}

Next, we derive the throughput of the frequency expansion scheme under anti-aliasing conditions, which is a similar process.  Returning to Equation \ref{eq:tp}, we must determine the limiting factor of the term $\min(\Delta f, f_0)$.  In this case, the smallest frequency spacing in the signal will be $\Delta f_X$, since $\Delta f_Y > \Delta f_X$ for this scheme.  And from the anti-aliasing analysis, the lowest frequency present in the output signal will be $(r_0+1)\Delta f_Y + (1-N)\Delta f_X$.  When plugging in the design choices of $\Delta f_Y = N \Delta f_X$ and $r_0 = 0$, the lowest frequency present simply becomes $\Delta f_X$.  Thus, the limiting factor of the term $\min(\Delta f, f_0)$ is $\Delta f_X$ for the frequency expansion scheme.

Hence, the throughput becomes $T_{\text{expansion}} = N \cdot R \cdot \Delta f_X = R \Delta f_Y$, using the design choice above for the second equality.  Dividing the throughput by the available bandwidth $B$, we get:

\begin{linenomath*}
\begin{align*}
\frac{T_\text{expansion}}{B} &= \frac{R \Delta f_Y}{ N \Delta f_X + (r_0 + R) \Delta f_Y} \\
&= \frac{R \Delta f_Y}{ N \left(\frac{1}{N}\Delta f_Y \right) + (0 + R) \Delta f_Y} \\
&= \frac{R \Delta f_Y}{(1+R) \Delta f_Y} \\
&= \frac{R}{1+R} \\
& \approx 1, \ \ \ \text{for} \ \ R \gg 1. \\
\end{align*}
\end{linenomath*}

Finally, the exact and approximate throughput of the frequency expansion scheme are:

\begin{linenomath*}
\begin{align}
T_\text{expansion} &= \frac{R}{1 + R} B \approx B
\label{eq:tpe}.
\end{align}
\end{linenomath*}

As discussed in the main text, to compute the convolutional throughput a multiplicative factor of $N$ is added to Equation \ref{eq:tp}, which results in directly applying that same factor to Equations \ref{eq:tpr} and \ref{eq:tpe}.  Note that in the main text, we assumed that the bandwidth $B$ is available not only for the input and weight signals, but also for the output signal.  So for the convolutional throughput, all the output spurious frequencies must fit within the bandwidth $B$.  Fortunately, we see from Equation \ref{eq:terms} that the highest frequency in the convolutional output is $(N-1) \Delta f_X + (r_0 + R) \Delta f_Y$, which is less than $B$.

\subsection{Non-WDM Throughput Derivation}

Here we derive the throughput of a single fully connected layer when not using optical WDM, meaning that it is limited to the electronic components.  Thus, in this case, the bandwidth $B$ is the bandwidth available from the DPMZMs, which was already derived in the previous section.  The other constraint is the photodetector with bandwidth $B_{PD}$.  The derivation for the throughput in terms of $B_{PD}$ is similar to that for $B$, except that in this case, the maximum available bandwidth is determined by the highest output neuron frequency, which yields:  $B_{PD} = (r_0 + R) \Delta f_Y$.

Starting with the frequency reduction scheme, we get:

\begin{linenomath*}
\begin{align*}
\frac{T_\text{reduction}}{B_{PD}} = \frac{N \Delta f_X}{(r_0 + R) \Delta f_Y}.
\end{align*}
\end{linenomath*}

Plugging in the anti-aliasing values of $\Delta f_Y = \frac{1}{R}\Delta f_X$ and $r_0 \Delta f_Y = \frac{1}{2}(N-\frac{R+1}{R})\Delta f_X$ and simplifying, we get:

\begin{linenomath*}
\begin{align*}
T_\text{reduction} &= \frac{2NR}{NR + R - 1} B_{PD} \approx 2 B_{PD}, \ \ \ \text{for} \ \ N \gg 1 
\end{align*}
\end{linenomath*}

And for the frequency expansion scheme, we have:

\begin{linenomath*}
\begin{align*}
\frac{T_\text{expansion}}{B_{PD}} &= \frac{R \Delta f_Y}{(r_0 + R) \Delta f_Y}
\end{align*}
\end{linenomath*}

Plugging in the anti-aliasing value $r_0 = 0$, we simply get:

\begin{linenomath*}
\begin{align*}
T_\text{expansion} = B_{PD},
\end{align*}
\end{linenomath*}

\noindent which is an exact value with no approximations.

Therefore for a given layer in the architecture, when limited to a modulation bandwidth of $B$ and a single photodetector with bandwidth $B_{PD}$, the maximum throughput is:

\begin{linenomath*}
\begin{align}
T_\text{reduction} & \approx \min(\frac{2}{3}B, \ 2 B_{PD}) \label{eq:redpd} \\
T_\text{expansion} & \approx \min(B, \ B_{PD}). \label{eq:exppd}
\end{align}
\end{linenomath*}

Additionally, Equations \ref{eq:redpd} and \ref{eq:exppd} show the relationship between the modulation bandwidth $B$ (the highest frequency among the input and weight signals) and the photodetector bandwidth $B_{PD}$ (the highest output neuron frequency).  Equation \ref{eq:exppd} implies that for large $R$ in the frequency expansion scheme, the maximum frequency contained among the input and weight signals is the same as the maximum frequency contained in the output neuron signal.  This relation between $B$ and $B_{PD}$ can be confirmed by comparing their definitions, and supports the intuition that the electronics-limited throughput is simply the minimum of the modulator and photodetector bandwidths.  And similarly for the frequency reduction scheme, Equation \ref{eq:redpd} implies that the maximum frequency among the input and weight signals is three times larger than that of the output neuron signal for large $N$.

\subsection{DNN Random Partial Sum Training}

Reading out the output vector after the photoelectric multiplication requires knowledge of the exact frequency content entering the system.  Additionally, analog filters are required to isolate the output vector frequencies in both the frequency reduction and expansion schemes.  However, in some applications it is desirable to have the flexibility to train a DNN anywhere in the spectrum without prior knowledge of the input, weight, and output frequencies.

\begin{table*}
\renewcommand\tabularxcolumn[1]{m{#1}}
\begin{tabularx}{0.8\textwidth} { 
  | >{\centering\arraybackslash}X 
  || >{\centering\arraybackslash}X 
  | >{\centering\arraybackslash}X 
  | >{\centering\arraybackslash}X |}
    \hline
    \textbf{DNN Shape} & \textbf{Conventional Accuracy} & \textbf{MAFT-ONN Accuracy} & \textbf{MAFT-ONN Accuracy (Partial Sum Training)}  \\
    \hline
    $7 \times 7 \to 32 \to 16 \to 10$ &  94.51\% & 93.97\% & 85.69\% \\
    \hline
    $14 \times 14 \to 32 \to 16 \to 10$  & 95.12\% & 95.98\% & 86.45\% \\
    \hline
    $28 \times 28 \to 1000 \to 1000 \to 10$ & 97.98\% & 97.93\% & 87.42\% \\
    \hline
    $7 \times 7 \to 100 \to 100 \to 100 \to 100 \to 100 \to 10$ & 97.05\% & 95.80\% & 93.52\% \\
    \hline
    $28 \times 28 \to 100 \to 100 \to 100 \to 100 \to 100 \to 10$ & 97.56\% & 96.35\% & 94.85\% \\
    \hline
\end{tabularx}
\caption{A comparison of the accuracy on the MNIST dataset on conventional DNNs with fully connected layers followed by a ReLU nonlinear activation versus simulations of the MAFT-ONN with the MZM nonlinear activation.}
\label{table:train}
\end{table*}

Recall from Supplementary Section G that the spurious frequencies contain unique partial sums.  Since these partial sums contain weight matrix elements, they too are trainable parameters for the DNN.  Thus, assigning the output vector to random frequencies in the spectrum will result in training the DNN not with a conventional fully connected layer, but instead with a random assortment of partial sums.  We call this method of DNN training \textit{random partial sum training}.

We explored the effects of random partial sum training on larger DNNs using simulations, shown in Table \ref{table:train}.  The hardware setup in these simulations is different than in the 3-layer experiment.  In the simulations, a bandpass filter is randomly placed in the spectrum within the general range of expected frequencies for each layer.

Table \ref{table:train} compares the DNN accuracy between a conventional DNN using fully connected layers with each followed by a ReLU nonlinear activation, a simulation of the MAFT-ONN with fully connected layers followed by the MZM nonlinearity, and a simulation of the MAFT-ONN with bandpass filters placed randomly in the spectrum. As shown in Table \ref{table:train}, the MAFT-ONN is on par with the accuracies of the conventional DNN.  However, the random partial sum training version of the MAFT-ONN loses accuracy for smaller workloads, but begins to approach the performance of the conventional DNN for larger workloads.  We hypothesize that the random partial sum training is less accurate because the density of partial sums in random places in the spectrum will be on average less than the density of partial sums at the output neuron values.  But as the size and depth of the DNN grows, the random partial sum training seems to converge to the conventional DNN accuracy, perhaps because the expressivity of the random partial sums increases.

In our 3-layer DNN experiment, we did not filter out the spurious frequencies after the first layer and randomly chose a set of output frequencies as the one-hot vectors for the MNIST classification.  Thus, we experimentally demonstrated a variant of partial sum training.

\subsection{Encoding for Complex Matrix Products}

Using complex encoding doubles the throughput of MAFT-ONN for a given amount of bandwidth $B$.  In addition, incoming waveforms in a real-world spectral environments such as radio transmissions will not only have an arbitrary amplitude, but also an arbitrary phase for each frequency mode.  Thus, we use a complex encoding scheme to reinterpret the amplitude and phase of each frequency mode into a complex number.  Here we show that the matrix operations are still achieved even for complex-valued elements.

In this case, the electrical voltage signal for the inputs includes both an amplitude $X_{n}$ and a phase $\phi^{X}_{n}$ for each frequency mode.  These can be interpreted as a complex number with real part $x^{\text{RE}}_n=X_n \cos(\phi^{X}_{n})$ and imaginary part $x^{\text{IM}}_n=-X_n \sin(\phi^{X}_{n})$.  Thus, the input activation is expressed as:

\begin{linenomath*}
\begin{align*}
&V_X(t) = \sum_{n=1}^{N}{X_{n} \cos((n_0 + n) \Delta \omega_X t + \phi^{X}_{n})} \\
&= \sum_{n=1}^{N}{x^{\text{RE}}_n \cos((n_0 + n) \Delta \omega_X t) + x^{\text{IM}}_n \sin((n_0 + n) \Delta \omega_X t)}.
\end{align*}
\end{linenomath*}

Similarly, each frequency mode in the weight matrix signal also has an amplitude $W_{r,n}$ and phase $\phi^{W}_{r,n}$, where we interpret it as a complex number with real part $w^{\text{RE}}_{r,n}=W_{r,n} \cos(\phi^{W}_{r,n})$ and imaginary part $w^{\text{RE}}_{r,n}=-W_{r,n} \sin(\phi^{W}_{r,n})$.  Then the weight matrix is expressed as:

\begin{linenomath*}
\begin{align*}
& V_W(t) = \sum_{r=1}^{R}\sum_{n=1}^{N}{W_{r,n} \cos(\omega_{r,n}^W t + \phi^{W}_{r,n})} \\
&= \sum_{r=1}^{R}\sum_{n=1}^{N}{w^{\text{RE}}_{r,n} \cos(\omega_{r,n}^W t) + w^{\text{IM}}_{r,n} \sin(\omega_{r,n}^W t)}.
\end{align*}
\end{linenomath*}

Following the same procedure as in the general MAFT algorithm, multiplying these inputs and weights yields the signal of the photoelectric output vector:

\begin{linenomath*}
\begin{align*}
V_Y(t) & \propto \sum_{r=1}^{R}{Y_{r} \sin(\omega_r^Y t + \phi^{Y}_{r})} \\
&\propto \sum_{r=1}^{R}{y^{\text{RE}}_n \sin(\omega_r^Y t) + y^{\text{IM}}_n \cos(\omega_r^Y t)},
\end{align*}
\end{linenomath*}

\noindent where $y^{\text{RE}}_r=Y_r \cos(\phi^{Y}_{r})$ and $y^{\text{IM}}_r=-Y_r \sin(\phi^{Y}_{r})$.

The expressions of $y^{\text{RE}}_r$ and $y^{\text{IM}}_r$ are determined by inserting the physical (amplitude and phase) representations of $V_X(t)$ and $V_W(t)$ into the general MAFT algorithm equations, then converting to the complex representation and simplifying.  This yields the complex representation of the output vector:

\begin{linenomath*}
\begin{align*}
y^{\text{RE}}_r &=  \sum_{n=1}^{N}{w^{\text{RE}}_{r,n} x^{\text{RE}}_n + w^{\text{IM}}_{r,n} x^{\text{IM}}_n} \\
&= \sum_{n=1}^{N}{W_{r,n} X_n \cos(\phi^{W}_{r,n} - \phi^{X}_{n})} \\
y^{\text{IM}}_r &=  \sum_{n=1}^{N}{w^{\text{RE}}_{r,n} x^{\text{IM}}_n - w^{\text{IM}}_{r,n} x^{\text{RE}}_n} \\
&= \sum_{n=1}^{N}{W_{r,n} X_n \sin(\phi^{W}_{r,n} - \phi^{X}_{n})}.
\end{align*}
\end{linenomath*}

And the physical representation of the output vector is:

\begin{linenomath*}
\begin{align*}
Y_r &=  \sqrt{ \left( y^{\text{RE}}_r  \right)^2 + \left( y^{\text{IM}}_r \right)^2 } \\
\phi^{Y}_{r} &= \tan^{-1} \left( \frac{- y^{\text{IM}}_r }{y^{\text{RE}}_r} \right).
\end{align*}
\end{linenomath*}

Here, we have described the physics of the MAFT-ONN when processing complex-valued signals, expressing each signal both in its physical and complex number representation.  Next, we show that this physical encoding corresponds to a complex-valued matrix-vector product.

Let us define a complex-valued input vector $X^{\mathbb{C}}$ with elements $X^{\mathbb{C}}_n = x^{\text{RE}}_n + i x^{\text{IM}}_n$.  Similarly we define a complex-valued weight matrix $W^{\mathbb{C}}$ with elements $W^{\mathbb{C}}_{r,n} = w^{\text{RE}}_{r,n} + i w^{\text{IM}}_{r,n}$.  Finally, we define the complex-valued output vector $Y^{\mathbb{C}} = \left( W^{\mathbb{C}} \right)^* X^{\mathbb{C}}$.  The elements of $Y^{\mathbb{C}}$ are:

\begin{linenomath*}
\begin{align*}
Y^{\mathbb{C}}_r &= \sum_{n=1}^{N}{\left( W^{\mathbb{C}}_{r,n} \right)^* X^{\mathbb{C}}_n} \\
&= \sum_{n=1}^{N}{\left( w^{\text{RE}}_{r,n} + i w^{\text{IM}}_{r,n} \right)^* \left( x^{\text{RE}}_n + i x^{\text{IM}}_n \right)} \\
&= \sum_{n=1}^{N}{w^{\text{RE}}_{r,n} x^{\text{RE}}_n + w^{\text{IM}}_{r,n} x^{\text{IM}}_n} \\
& \ \ \ \ + i \sum_{n=1}^{N}{w^{\text{RE}}_{r,n} x^{\text{IM}}_n - w^{\text{IM}}_{r,n} x^{\text{RE}}_n}
\end{align*}
\end{linenomath*}

The last expression matches that of the physical encoding, yielding $Y^{\mathbb{C}}_r = y^{\text{RE}}_r + i y^{\text{IM}}_r$.  Therefore, the physical output of the MAFT-ONN corresponds to a complex matrix-vector product.

\begin{figure*}[t]
\centering
\includegraphics[width=\textwidth]{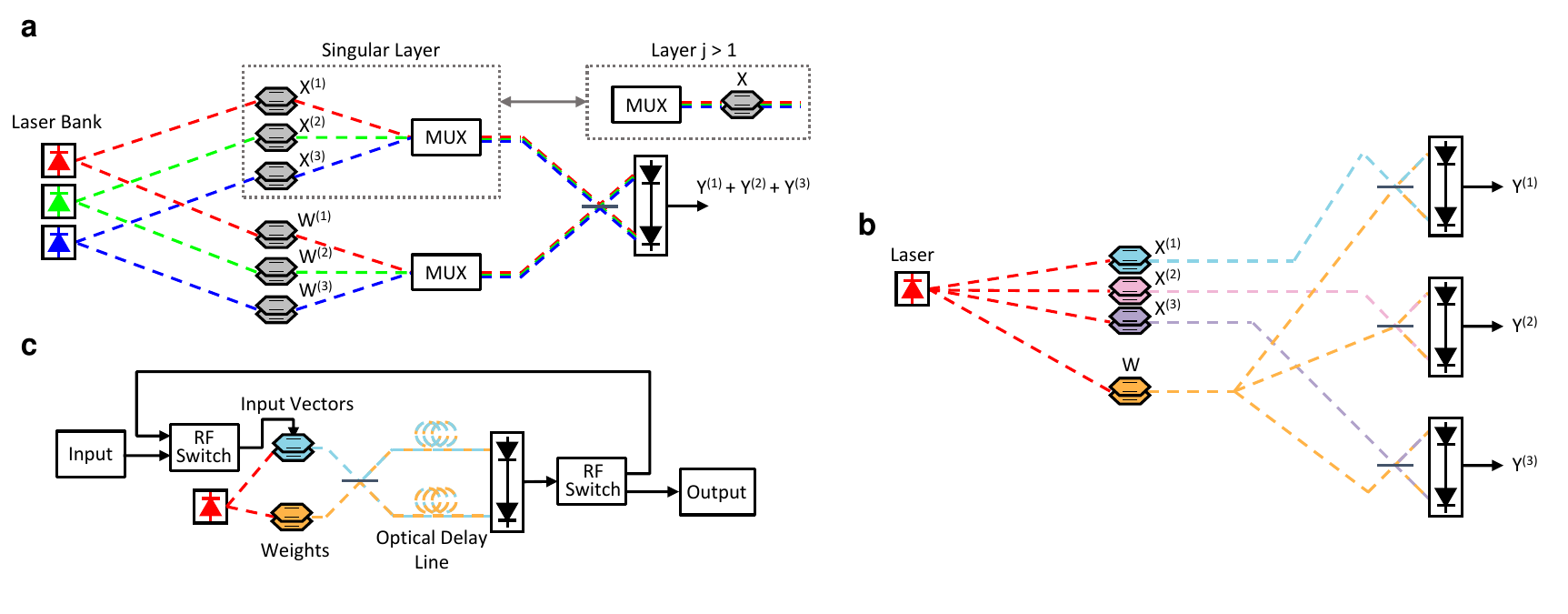}
\caption{(a)  A WDM version of this architecture that expands the bandwidth-limited throughput to the wide bandwidth available in optics.  Several matrix-vector products can be independently computed on different laser wavelengths, where the WDM output incoherently sums of the independent output vector signals from each laser.  (b)  A spatially multiplexed version of this architecture that uses optical fan-out to reuse the weight matrix for multiple input vectors.  This variant can increase both throughput and energy efficiency.  (c)  The ``loop'' version of the MAFT-ONN that can implement an arbitrary number of layers with a signal set of modulators.  A fiber delay line can be used to allow time for the RF weight values to update for each layer and for the RF switches to route the data.}
\label{fig:fut}
\end{figure*}

\subsection{Maximizing Throughput with Scalable Multiplexing}

The computational throughput discussion claimed that the bandwidth $B$ of the throughput is not limited by the electro-optic RF components but instead by the available optical bandwidth.  Hence, maximum throughput can be achieved by maximizing the use of the optical spectrum.  Here we present scalable architectures that allow MAFT-ONN to increase throughput by taking advantage of optics.

\noindent \textbf{Scalability in DNN Width:}  The ''width" of a DNN refers to the number of weights that can fit into a single layer.  This is directly proportional to the amount of optical bandwidth being utilizes.  As mentioned in the main text, the two primary methods to utilize optical bandwidth with MAFT-ONN are (i) optical wavelength-division multiplexing (WDM) the frequency-encoded signals, or (ii) replacing the frequency-encoded signals with optical frequency combs.  For (i),  Figure \ref{fig:fut}(a) illustrates a variant of this architecture that uses optical WDM to simultaneously perform multiple matrix-vector products on each laser on the same photodetector.  The incoherence between the lasers allows for each matrix-vector product to independently sum at the photovoltage output.  As long as the gap between each laser wavelength is greater than the bandwidth of the photodetector, there will be no cross coupling terms between the matrix-vector products performed on each laser carrier.  With the WDM version of the architecture, large matrix products can be tiled in the frequency domain, or matrix-matrix products can be frequency-multiplexed while still computing everything in a single shot.  The optical bandwidth can also be used in the case of an arbitrarily deep neural network (the box labeled ``Layer $j>1$'' in Figure \ref{fig:fut}(a)), where the same input vector can be independently multiplied by different weight signals for applications like convolution.  There is more than 20 THz of available bandwidth among the S, C, and L telecommunication bands (1460nm-1625nm) that can be used here for optical WDM.  For (ii), optical frequency combs can replace the WDM modules in Figure \ref{fig:fut}(a) for even larger throughput.  Other works have experimentally demonstrated optical frequency combs with almost 1,000 THz bandwidth \cite{lesko2021six, couny2007generation}.  An optical AWG or waveshaper would be required to program the frequency combs.
 
The throughput can be further increased through spatial multiplexing with copies of the setup running in parallel, shown in Figure \ref{fig:fut}(b).  Demonstrations like a photonic integrated circuit with 48 on-chip MZMs \cite{streshinsky2014silicon} show the promise for the scalability of MAFT-ONN.

Note that this throughput analysis assumes that the optical bandwidth $B$ is the limiting factor in throughput.  See Supplementary Section J for a throughput derivation that is limited to only the electronics without using optical multiplexing.

\noindent \textbf{Scalability in DNN Depth:}  The ``depth'' of a DNN concerns the number of layers.  Another multiplexing variant of MAFT-ONN is illustrated in Figure \ref{fig:fut}(d) that can compute an arbitrary number of DNN layers with a single set of modulators.  This ``loop'' version uses an optical fiber delay lines as temporary optical storage to give time for the RF weights and data routing switches to operate.  Just 1 km of commercially available optical fiber used as a delay line is enough to enable MHz-speed RF switches.  This version of the architecture can reduce the cost, hardware complexity, and power consumption for computing DNNs with many hidden layers.  The primary limiting factor here is determining how many DNN layers can be implemented before the noise degrades the performance.

\subsection{Frequency-Domain LTI Framework}

Here we show how the physics of MAFT-ONN correspond to LTI operations.  Let the electrical input and LTI filter signals be $V_X(t)$ and $V_W(t)$ respectively.  Assuming both signals are frequency-encoded with the same frequency spacing $\Delta \omega$ yields: $V_X(t)=\sum_{n=1}^{N}{X_n \cos(n \cdot \Delta \omega \cdot t)}$ and $V_W(t)=\sum_{r=1}^{R}{W_r \cos(r \cdot \Delta \omega \cdot t)}$. 

As in the main text, we program the DPMZM biases such that $E_X(t)$ is SSB-SC with the right sideband suppressed and $E_W(t)$ is SSB-SC with the left sideband suppressed.  Assuming both signals are frequency-encoded with the same frequency spacing $\Delta \omega$, this yields:

\begin{linenomath}
\begin{align}
E_X(t) \propto \sum_{n=1}^{N}{X_{n} e^{i(-n \Delta \omega + \omega_{\text{LD}}) t}}, \label{eq:ex}\\
E_W(t) \propto \sum_{r=1}^{R}{W_{r} e^{i(r \Delta \omega + \omega_{\text{LD}}) t}}. \label{eq:ew}
\end{align}
\end{linenomath}

The photoelectric multiplication then yields the output voltage: $V_{\text{out}}(t) \propto \text{Im} \left[E_X^*(t)E_W(t)\right]$.  Taking the Fourier Transform of both sides yields:

\begin{linenomath}
\begin{align}
\mathcal{F} \left[ V_{\text{out}}(t) \right] \propto \left(\Tilde{E}_X \star \Tilde{E}_W \right)(\omega) - \left(\Tilde{E}_W \star \Tilde{E}_X \right)(\omega) \label{eq:cross}
\end{align}
\end{linenomath}

\noindent where the star indicates the cross correlation function, $\Tilde{E}_X(\omega) = \mathcal{F} \left[ E_X(t) \right]$ and $\Tilde{E}_W(\omega) = \mathcal{F} \left[ E_W(t) \right]$.  The continuous cross correlation between $\Tilde{E}_X(\omega)$ and $\Tilde{E}_W(\omega)$ is defined as:

\begin{linenomath}
\begin{align}
\left(\Tilde{E}_X \star \Tilde{E}_W \right)(\omega) = \int_{-\infty}^{\infty}{\Tilde{E}_X^*(\theta)}\Tilde{E}_W(\omega + \theta) \  d \theta \label{eq:integral}
\end{align}
\end{linenomath}

Equations \ref{eq:ex} through \ref{eq:integral} can be combined and simplified to yield:

\begin{linenomath}
\begin{align}
\mathcal{F} \left[ V_{\text{out}}(t) \right] \propto \sum_{r=1}^{R}\sum_{n=1}^{N}{X^*_n W_x \delta (\omega - (x + n) \Delta \omega)} \nonumber \\  
- X_n W^*_x \delta (\omega + (x + n) \Delta \omega) \label{eq:crossfin}.
\end{align}
\end{linenomath}

\noindent where $\delta(\cdot)$ is the Dirac delta function.

Note that instead of the frequencies subtracting like in Supplementary Section G, they add here when the DPMZM biases are programmed to place $E_X(t)$ and $E_W(t)$ on opposite sidebands.  The terms in the summation containing $X^*_n W_x$ strictly correspond to the positive frequencies and the terms with $X_n W^*_x$ strictly correspond to the negative frequencies.  Hence there is no aliasing between the two terms in the summation.  And when taking the Inverse Fourier Transform of Equation \ref{eq:crossfin}, a summation of real-valued sine waves with various phases emerges.

Due to the lack of positive/negative frequency aliasing between the terms in Equation \ref{eq:crossfin} we can just consider the terms that correspond to the positive frequencies for the LTI analysis:

\begin{linenomath}
\begin{align}
\mathcal{F} \left[ V_{\text{out}}(t) \right]_+ \propto \sum_{r=1}^{R}\sum_{n=1}^{N}{X^*_n W_x \delta (\omega - (x + n) \Delta \omega)} \label{eq:crosspos}.
\end{align}
\end{linenomath}

Now we define the frequency-domain LTI signals and show that the LTI convolution operator matches the physics of MAFT-ONN.  We first define a simple LTI scheme were we allow information to be encoded at DC.  Let $x[n] \equiv X_n \rightarrow n \Delta \omega$ and $w[n] \equiv W_n \rightarrow n \Delta \omega$.  Hence the values of $x[n]$ and $w[n]$ are mapped to the frequency mode at $n \Delta \omega$.  Thus the LTI signals are:

\begin{linenomath}
\begin{align*}
x[n] = \sum_{m=1}^{N}{X_m \delta [n - m]} \\
w[n] = \sum_{m'=1}^{R}{W_{m'} \delta [n - m']}.
\end{align*}
\end{linenomath}

\noindent where $\delta[\cdot]$ is the Kronecker delta function and $m'$ is distinguished from $m$ for indexing convenience.

Then the convolution operator $(x * w)[n]$ yields:

\begin{linenomath}
\begin{align}
(x * w)[n] &= \sum_{k=-\infty}^{\infty}{x[k] w[n - k]} \nonumber \\ 
&= \sum_{m'=1}^{R}\sum_{m=1}^{N}{X_m W_{m'} \delta [n - (m' + m)]} \label{eq:lti0}.
\end{align}
\end{linenomath}

Therefore when considering real values for $x[n]$ and $w[n]$, the LTI convolution in Equation \ref{eq:lti0} matches the MAFT-ONN physics in Equation \ref{eq:crosspos}.  When considering complex values we still achieve a useful LTI-like convolution.

However, this LTI scheme requires programming the information of $x[0]$ and $w[0]$ at DC.  To avoid this we use the LTI framework in the main text where we offset the indexing by 1 such that $x[n] \equiv X_{n+1} \rightarrow (n + 1) \Delta \omega$ and $w[n] \equiv W_{n+1} \rightarrow (n+1) \Delta \omega$.  Now the values $x[n]$ and $w[n]$ correspond to the frequency mode at $(n+1) \Delta \omega$.  The new LTI signals are:

\begin{linenomath}
\begin{align*}
x[n] = \sum_{m=1}^{N}{X_m \delta [n + 1 - m]} \\
w[n] = \sum_{m'=1}^{R}{W_{m'} \delta [n + 1 - m']}.
\end{align*}
\end{linenomath}

And now the convolution operator $(x * w)[n]$ yields:

\begin{linenomath}
\begin{align}
(x * w)[n] &= \sum_{k=-\infty}^{\infty}{x[k] w[n - k]} \nonumber \\ 
&= \sum_{m'=1}^{R}\sum_{m=1}^{N}{X_m W_{m'} \delta [n + 2 - (m' + m)]} \label{eq:lti1}.
\end{align}
\end{linenomath}

With this framework, the output LTI signal $y[n] = (x * w)[n]$ can be physically interpreted similarly as $x[n]$ and $w[n]$ where the value of $y[n]$ corresponds to the frequency mode at $(n+1) \Delta \omega$.  This keeps the frequency-domain interpretation of $x[n]$, $w[n]$, and $y[n]$ consistent while avoiding programming information at DC.

However, note that with this offset LTI framework, comparing Equation \ref{eq:lti1} with the physics of MAFT-ONN in Equation \ref{eq:crosspos} reveals that MAFT-ONN actually physically computes:  $y_{\text{MAFT}}[n] = y[n-1] = (x * w)[n] * \delta [n-1]$.  This means that MAFT-ONN physically pushes $y[n]$ up by one $\Delta \omega$ for every LTI convolution.  For the overwhelming majority of use-cases this is non-problematic as long as one properly keeps track of the frequencies.  But if one truly wishes, the extra $\delta [n-1]$ can be removed with another layer of MAFT by multiplying $y_{\text{MAFT}}[n]$ with a single tone using same-sideband DPMZM biasing.  Because LTI operations are cascadable and the order does not matter, this can be done at the very end of a chain of MAFT-ONN LTI convolutions.

Furthermore, the \textit{linearity} and \textit{time invariance} requirements of this frequency-domain LTI framework are clearly upheld by the physics of MAFT-ONN in Equation \ref{eq:crosspos}, which has been validated by the numerous experiments in this work.  The MAFT-ONN linearity is upheld as long as the DPMZMs operate in their linear regime and the time invariance is upheld from the fact that the same LTI convolution is computed regardless of where in the spectrum the signal lies.  Due to how we mapped the LTI signals to the physics, all signals anywhere in the MAFT-ONN frequency domain LTI framework will have zero magnitude for $n<0$.

Note that the Dirac delta functions in Equation \ref{eq:crosspos} can be pulse-shaped using a frequency communications scheme like orthogonal frequency division multiplexing (OFDM) and still essentially retain the LTI functionality albeit with finite resolution.

\subsection{Communication Link Gain Analysis}

The gain of single layer of MAFT-ONN is expressed below \cite{urick2015fundamentals}:

\begin{linenomath}
\begin{align}
g \ (\text{linear}) = \frac{\pi^2}{8} \left(\frac{R_{PD} \gamma P_{\text{LD}}}{V_{\pi}} \right)^2 R_i R_o \left |{H_{PD}(f)}\right|^2 \left \langle V_W^2(t) \right \rangle
\label{eq:g}
\end{align}
\end{linenomath}

where $R_{PD}$ is the responsivity of the photodetector, $\gamma$ is the gain of the optical link (modulator insertion loss, fiber propagation loss, optical amplifiers, etc.), $P_{LD}$ is the laser power, $V_{\pi}$ is the voltage required to reach $\pi$ phase shift on the modulators, $R_i$ and $R_o$ are the input and output resistances respectively, $H_{PD}(f)$ is the frequency response of the photodetector, and $\left \langle V_W^2(t) \right \rangle$ is the time-averaged power of the weight matrix signal.  Note that this equation is for a receiverless link (no RF amplifier following the balanced photodetector).

\begin{figure}[h]
\includegraphics[width=0.9\linewidth]{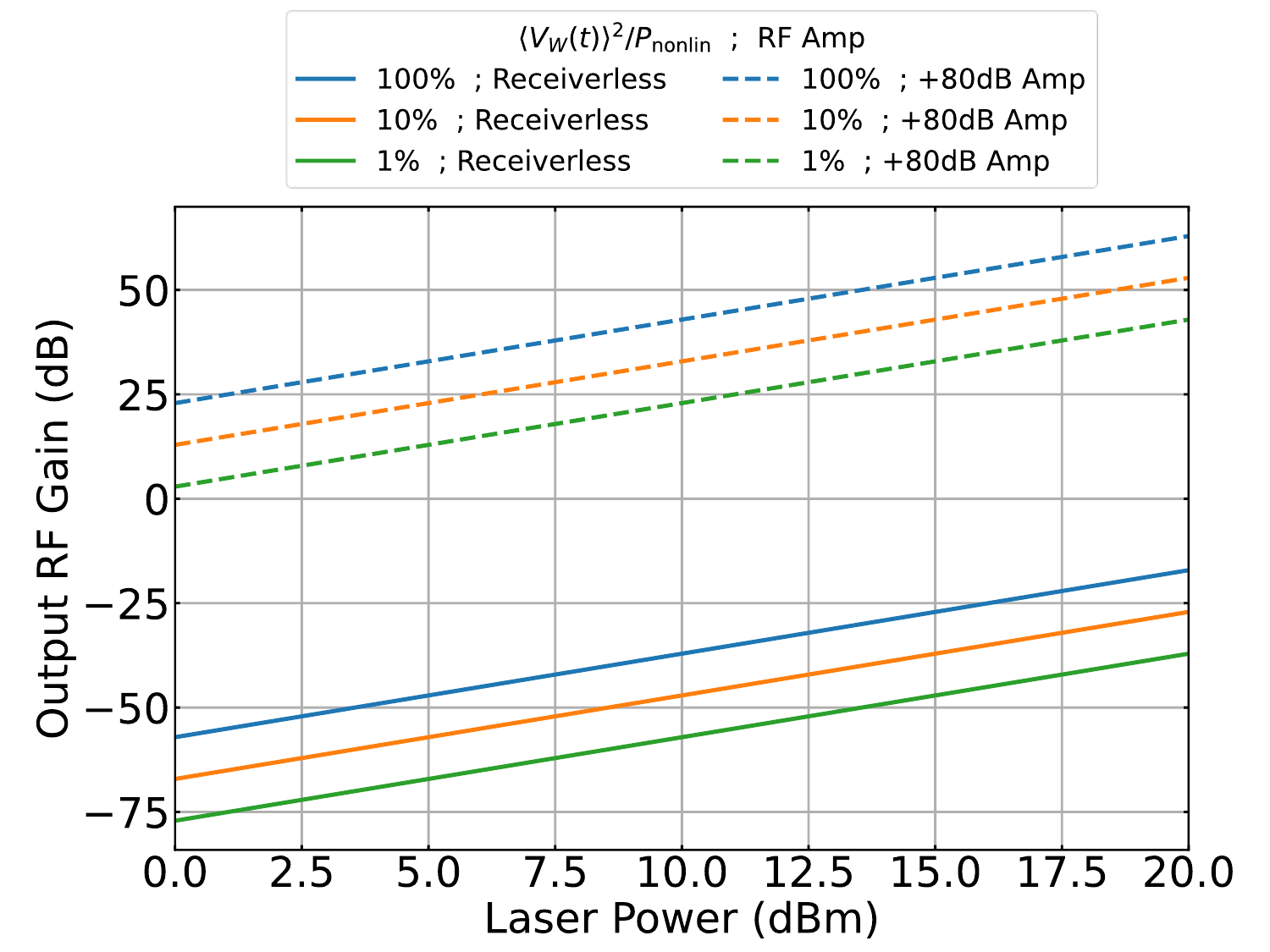}
\caption{A communication link gain analysis, illustrating the trade-space between the laser power, weight signal power, and RF amplifier gain from Equation \ref{eq:g}.  The weight signal can be any power as long as it stays within the linear regime of the modulator, where nonlinear power threshold of the modulator is $P_{\text{nonlin}}$.} \label{fig:gain}
\end{figure}

Figure \ref{fig:gain} illustrates a trade-space between the laser power, the weight signal power, and an RF amplifier.  In the plot, $V_{\pi}$ is 6 V,$R_{PD}$ is 1 A/W, $\gamma$ is -6 dB, $R_i$ and $R_o$ are $50 \ \Omega$, and $H_{PD}$ is 1/2.  Since the power of the weight signal can be adjusted to fit within the linear regime of any modulator, the gain curves are independent of the $V_{\pi}$ of the modulators and instead depend on $\left \langle V_W^2(t) \right \rangle$.  However, the $V_{\pi}$ will still determine the threshold of nonlinear regime of the modulator implementing the nonlinear activation.

In our experiment, we found that our DPMZM with $V_{\pi} \approx 6 \ V$ did not exhibit nonlinear behavior until the input signal reached around $P_{\text{nonlin}} = V_{\pi}^2 / R_i \approx 27 \ \text{dBm}$.  To alleviate the requirement of high power signals to reach the nonlinear regime, note that in principle, modulators with $V_{\pi} \approx 1 \ \text{mV}$ can be fabricated \cite{hochberg2007towards}.  In that case, the nonlinear power threshold of the modulator is $P_{\text{nonlin}} \approx -47 \ \text{dBm}$.  This even allows for input RF signals with -85 dBm of power, which is typically considered the minimum usable power level for communications \cite{bhargav2022prediction}, to be amplified enough to reach the nonlinear regime.  Therefore in some scenarios, the gain from the laser may allow for receiverless operation, and in others an amplifier either before or after the modulators may be required.

\subsection{Latency Estimation Model}

Here we estimate the theoretical fastest latency of various digital processors for the task of applying a filter to an incoming wireless signal with bandwidth $B$.  The filter is a simple 1-dimensional convolution in the frequency-domain with frequency resolution $\Delta f$.  Hence for a signal with bandwidth B and frequency resolution $\Delta f$, the Nyquist sampling theorem requires at least $N=2 \lceil \frac{B}{\Delta f} \rceil$  points to digitally capture this signal.  Finally, the filtered signal is moved to a device that can implement the next stage of processing as necessary.

Note that these are models of electronics working at their idealized peak speeds, using optimistic parameters (compared to what is used in practice), with 100\% processor and memory utilization, without the additional processing management overhead, assuming no data rate degradation from high clock speeds, and without throttling due to overheating.  Hence, we are comparing MAFT to the theoretical limits of electronics that are usually not even achieved in practice.

Figure 5(a) in the main text shows the various components modeled here that contribute to the latency of the various computing architectures.  Note that GPUs will likely suffer the most additional overhead that cannot be deterministically predicted by a model, which influences the latencies achieved in practice.

\begin{table*}
\renewcommand\tabularxcolumn[1]{m{#1}}
\begin{tabularx}{0.8\textwidth} { 
  | >{\centering\arraybackslash}X 
  || >{\centering\arraybackslash}X 
  | >{\centering\arraybackslash}X 
  | >{\centering\arraybackslash}X |}
    \hline
    \textbf{GPU Model} & \textbf{$r_{\text{clock}}$} & \textbf{$n_{\text{PE}}$} & \textbf{$r_{\text{VRAM}}$}  \\
    \hline
    GeForce RTX 4090 &  2253 MHz (2520 MHz boosted) & 16,384 CUDA cores & 1,008 GB/s \\
    \hline
    RTX 5000 Ada Generation &  1155 MHz (2550 MHz boosted) & 12,800 CUDA cores & 576 GB/s \\
    \hline
\end{tabularx}
\caption{The parameters and devices used in the latency estimation model for the GPU RF receiver architecture.  All our GPU latency estimations use the boosted clock frequencies.}
\label{table:gpu}
\end{table*}

\noindent \textbf{GPU Latency:}  The latency of filtering the RF signal using a typical GPU system is estimated to be:

\begin{linenomath}
\begin{align*}
\tau_{\text{GPU}} = \tau_{\text{Gabor}} + \tau_{\text{JESD}} + \tau_{\text{PCIe}} + \tau_{\text{VRAM}} \\ + \tau_{\text{MAC}} + \tau_{\text{VRAM}}
\end{align*}
\end{linenomath}

The first term $\tau_{\text{Gabor}}$  is the Gabor limit, which is a fundamental constraint between time and frequency resolution.  In order to achieve frequency resolution $\Delta f$, the signal must be sampled for $\tau_{\text{Gabor}} = 1 / \Delta f$ seconds.  We assume that the physical latency of the ADC is negligible compared to the Gabor limit.  For simplicity we assume that the ADC sampling rate is exactly double the RF bandwidth, thus sampling at a rate of 2B samples/second to produce 2N samples.  (The factor of two on the number of points derives from sampling both the I and Q components of the RF signal.)  To trade off between speed and accuracy, we assume each sample has 16-bit precision ($b$ = 2 bytes), thus producing $D=2Nb$ bytes of data.

The second term $\tau_{\text{JESD}}$  is the latency due to data transfer from the ADC to the CPU.  Here we assume the data rate is the maximum possible $r_{\text{JESDC}}$ = 32 Gb/s = 4 GB/s for JESD204C and $J_{\text{lanes}}$ = 16 parallel lanes.  Hence, this latency is $tau_{\text{JESD}} = \frac{D}{J_{\text{lanes}} r_{\text{JESDC}}}$.

The third term $\tau_{\text{PCIe}}$  is the latency due to data transfer to and from the CPU to the GPU, typically via PCIe.  Here we assume the maximum possible data rate of PCIe 4.0 of $r_{\text{PCIe4r}}$ = 2 GB/s with $P_{\text{lanes}}$ = 16 lanes to yield a latency of $\tau_{\text{PCIe}}  = \frac{D}{P_{\text{lanes}} r_{\text{PCIe4}}}$.

The fourth and sixth terms $\tau_{\text{VRAM}}$  is the time it takes for the GPU to transfer data from its VRAM to its compute.  This term depends on the specific memory bandwidth of the GPU $r_{\text{VRAM}}$, and is thus $\tau_{\text{VRAM}} = \frac{D}{r_{\text{VRAM}}}$.

The fifth term $\tau_{\text{MAC}}$ is the time it takes to compute the RF filter.  This is approximated by the expression:   $\tau_{\text{MAC}} = \frac{O(N)}{r_{\text{clock}} n_{\text{PE}}} = \frac{2N}{r_{\text{clock}} n_{\text{PE}}}$, where $O(N)$ is the number of MACs for the computation, the factor of 2 comes from the fact that the RF signal is complex, $r_{\text{clock}}$ is the clock speed of the GPU and $n_{\text{PE}}$ is the number of parallel processing cores. The clock speed and number of processing cores depends on the GPU.

\begin{table*}
\renewcommand\tabularxcolumn[1]{m{#1}}
\begin{tabularx}{0.8\textwidth} { 
  | >{\centering\arraybackslash}X 
  || >{\centering\arraybackslash}X 
  | >{\centering\arraybackslash}X 
  | >{\centering\arraybackslash}X 
  | >{\centering\arraybackslash}X |}
    \hline
    \textbf{FPGA Model} & \textbf{$n_{\text{PE}}$} & \textbf{$r_{\text{RAM}}$} & \textbf{$M_{\text{BRAM}}$} & \textbf{$M_{\text{URAM}}$}  \\
    \hline
    Versal AI Edge VE2802 &  1312 DSP cores & 102.4 GB/s & 21.1 Mb & 74.3 Mb \\
    \hline
    Versal HBM Series VH1782 &  10848 DSP cores & 819.2 GB/s & 132 Mb & 541 Mb \\
    \hline
\end{tabularx}
\caption{The parameters and devices used in the latency estimation model for the FPGA RF receiver architecture.  Note that we assume $r_{\text{clock}}$ = 500 MHz for our FPGA estimations.}
\label{table:fpga}
\end{table*}

\begin{figure*}[t]
\centering
\includegraphics[width=\textwidth]{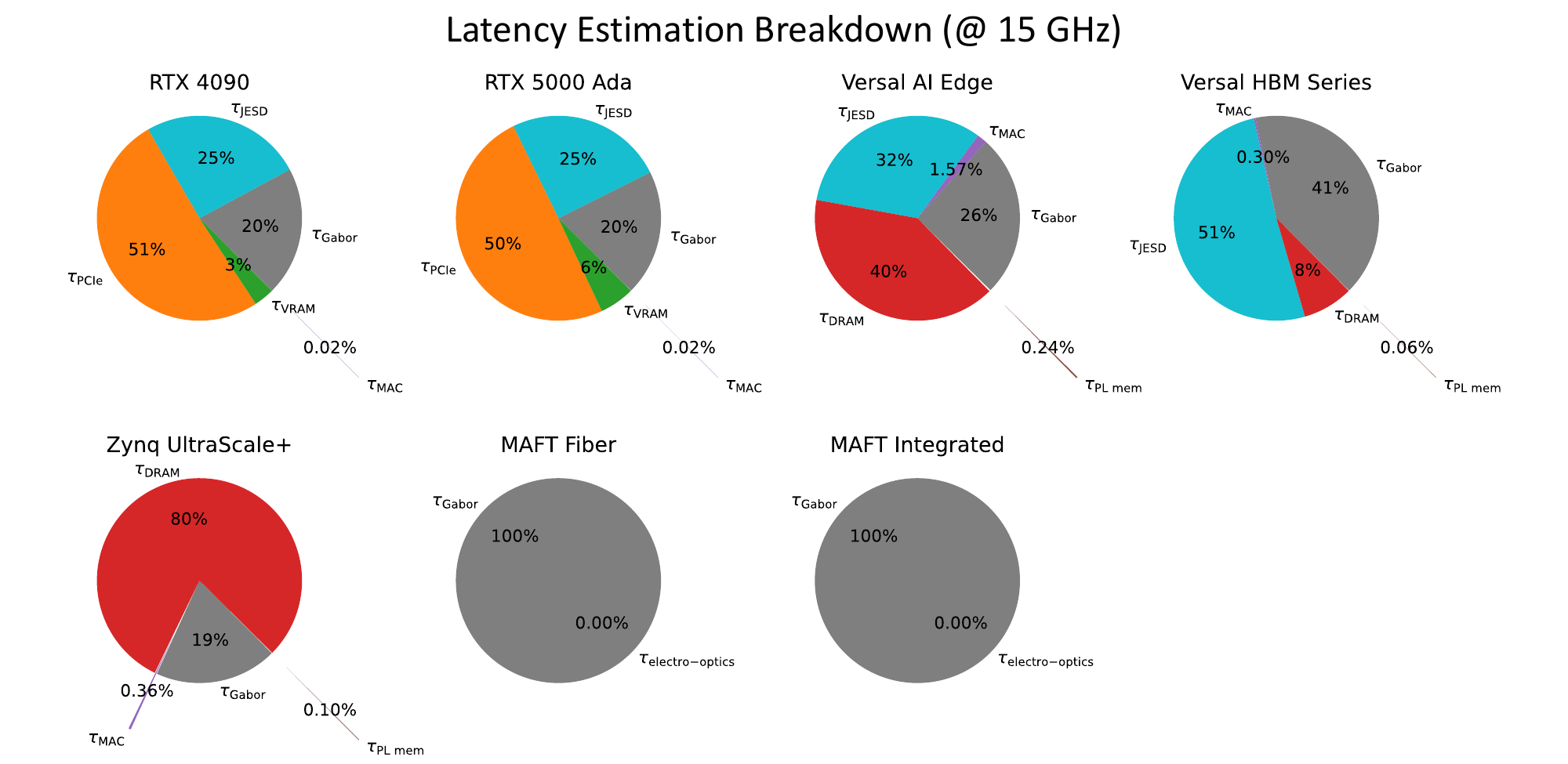}
\caption{A breakdown of the contributions to the total latency for the different evaluated RF receiver architectures evaluated at $B = 15$ GHz of input RF bandwidth.  The breakdown varies depending on $B$, where for lower values of $B$ the Gabor limit dominates and for higher values of $B$ the data movement dominates.}
\label{fig:bdown}
\end{figure*}

\noindent \textbf{FPGA Latency:} For digital processors, there is a tradeoff between the size of the RAM and the amount of overhead for data management.  FPGAs have significantly less RAM than GPUs but in return can directly process specialized operations with minimal overhead. The estimated ideal latency of an equivalent FPGA RF filter is:

\begin{linenomath}
\begin{align*}
\tau_{\text{FPGA}} = \tau_{\text{Gabor}} + \tau_{\text{JESD}} + \tau_{\text{RAM}} + \tau_{\text{PL mem}} \\ + \tau_{\text{MAC}} + \tau_{\text{RAM}}
\end{align*}
\end{linenomath}

The terms $\tau_{\text{Gabor}}$ and $\tau_{\text{JESD}}$ are the same as in the GPU model.

The third and seventh term $\tau_{\text{RAM}}$ is the latency due to data movement to/from the DRAM/HBM and the BRAM/URAM.  This is dominated by the maximum theoretical DRAM/HBM rate $r_{\text{RAM}}$ and the number of buses to the DRAM/HBM $R_{\text{lanes}}$, both of which depend on the specific FPGA model and setup.  The term $\tau_{\text{read}}$ accounts for the latency of initializing each memory read.  (Thus, we capture the penalty for having small programmable logic (PL) RAM.  Otherwise this model would not distinguish between an FPGA with a kilobyte versus a gigabyte of PL RAM.)  In memory architectures like DDR4, the latency for each memory call is dominated by the CAS latency, which is usually given in units of number of clock cycles.  Because there are various types of memory architectures we simplify the analysis by estimating the optimal read latency $n_{\text{read}}$ as 10 cycles for DRAM and HBM and 2 cycles for BRAM and URAM.  Thus:  $\tau_{\text{read}} =  \lceil \frac{D}{M} \rceil \frac{n_{\text{read}}}{r_{\text{clock}}}$ where M is the total PL RAM available to the FPGA and $r_{\text{clock}}$ is the clock speed of the FPGA (for simplicity we assume the clock shared for the DSP blocks and the RAM).  Therefore:  $\tau_{\text{RAM}} = \frac{D}{R_{\text{lanes}} r_{\text{RAM}}}$.

The fourth and sixth term $\tau_{\text{PL mem}}$ is the latency for transferring data to/from the PL memory (including BRAM and URAM) to the DSP blocks.  For simplicity we assume the bottleneck here is which is determined by minimum of the total BRAM and URAM memory bandwidths, $r_{\text{BRAM}}$ and $r_{\text{URAM}}$.  Because FPGA spec sheets usually only specify the amount of BRAM and URAM in Megabytes, we can estimate the memory bandwidth by using the basic architecture of each RAM.  Each BRAM block contains 36 kb with 2 36 bit ports, and each URAM block contains 288 kb with 2 72 bit ports.  Hence for a given amount of BRAM memory $m_{\text{BRAM}}$ and URAM memory $M_{\text{URAM}}$, the memory bandwidths can be estimated as $r_{\text{BRAM}} = 2 \cdot 36 \text{bits} \cdot  \lceil \frac{M_{\text{BRAM}}}{36 \text{kb}} \rceil r_{\text{clock}}$ and $r_{\text{URAM}} = 2 \cdot 72 \text{bits} \cdot \lceil \frac{M_{\text{URAM}}}{288 \text{kb}} \rceil r_{\text{clock}}$.

Therefore: $\tau_{\text{PL mem}} = \frac{1}{2} \left(1 / r_{\text{BRAM}} + 1 / r_{\text{URAM}} \right) D + \tau_{\text{read}}$.  Note that for simplicity we assume an even split in using BRAM and URAM, and the $\tau_{\text{read}}$ here depends on the read latency of BRAM/URAM instead of DRAM/HBM.

The term $\tau_{\text{MAC}}$ is the computational time it takes for the DSP blocks to compute the filter.  Similarly to the GPU case, this is estimated as:  $\tau_{\text{MAC}} = \frac{O(N)}{r_{\text{clock}} n_{\text{DSP}}} = \frac{2N}{r_{\text{clock}} n_{\text{DSP}}}$, where $r_{\text{clock}}$ is the FPGA clock frequency and $n_{\text{DSP}}$ is the number of DSP blocks on the FPGA, which depends on the model.

\begin{table*}
\renewcommand\tabularxcolumn[1]{m{#1}}
\begin{tabularx}{0.8\textwidth} { 
  | >{\centering\arraybackslash}X 
  || >{\centering\arraybackslash}X 
  | >{\centering\arraybackslash}X 
  | >{\centering\arraybackslash}X 
  | >{\centering\arraybackslash}X |}
    \hline
    \textbf{RF SoC Model} & \textbf{$n_{\text{PE}}$} & \textbf{$r_{\text{RAM}}$} & \textbf{$M_{\text{BRAM}}$} & \textbf{$M_{\text{URAM}}$}  \\
    \hline
    Zynq UltraScale+ RFSoC ZU49DR &  4272 DSP cores & 19.2 GB/s (DDR4 64-bit) & 38.0 Mb & 22.5 Mb \\
    \hline
\end{tabularx}
\caption{The parameters and devices used in the latency estimation model for the ADC-integrated FPGA RF SoC receiver architecture.  Note that we assume $r_{\text{clock}}$ = 500 MHz for these latency estimations.}
\label{table:rfsoc}
\end{table*}

\noindent \textbf{ADC-Integrated RF SoC Latency:}  Finally, the highest-performing ultra low-latency RF signal processing architecture are FPGAs with integrated ADCs and RF SoCs.  This eliminates the bottleneck of the data transfer between the ADC and the FPGA PL memory by directly linking them together.  The estimated ideal latency of this architecture is:

\begin{linenomath}
\begin{align*}
\tau_{\text{RFSoC}} = \tau_{\text{Gabor}} + \tau_{\text{PL mem}} + \tau_{\text{RAM}}
\end{align*}
\end{linenomath}

All of the terms here have already been defined in the previous sections.

\begin{table*}
\renewcommand\tabularxcolumn[1]{m{#1}}
\begin{tabularx}{0.8\textwidth} { 
  | >{\centering\arraybackslash}X 
  || >{\centering\arraybackslash}X 
  | >{\centering\arraybackslash}X 
  | >{\centering\arraybackslash}X |}
    \hline
    \textbf{MAFT-ONN Variant} & \textbf{$\tau_{\text{MZM}}$} & \textbf{$\tau_{\text{prop}}$} & \textbf{$\tau_{\text{PD}}$}  \\
    \hline
    MAFT (Untrimmed fiber) &  25 ps & 35 ns & 10 ps \\
    \hline
    MAFT (Integrated) &  25 ps & 30 ps & 10 ps \\
    \hline
\end{tabularx}
\caption{The parameters used in the latency estimation model for the MAFT variants of the RF receiver architecture.}
\label{table:maft}
\end{table*}

\noindent \textbf{MAFT Latency:}  The latency of MAFT with $J$ cascaded layers is:

\begin{linenomath}
\begin{align*}
\tau_{\text{MAFT}} = J(\tau_{MZM} + \tau_{PD} + \tau_{RF} + \tau_{\text{prop}}) + \tau_{\text{Gabor}},
\end{align*}
\end{linenomath}

\noindent where $\tau_{MZM}$ is the reciprocal of the bandwidth of the MZM, $\tau_{PD}$ is the reciprocal of the bandwidth of the photodetector, $\tau_{RF}$ is the combined delay due to the bandwidth of additional RF components like a bandpass filter or amplifier, and $\tau_{\text{prop}}$ is the data movement in the form of propagation of the frequency-encoded electromagnetic waves.

The value of $\tau_{MZM}$ highly depends on the material used for the MZM.  State-of-the-art commercial MZMs typically have up to 40 GHz bandwidth, contributing $\sim 25$ ps delay.  The photodetector latency can be separated into the RC time constant and carrier transit time: $\tau_{PD} = \sqrt{\tau_{RC}^2 + \tau_{transit}^2}$ \cite{chrostowski_hochberg_2015}.  Whether the RC or carrier transit time will dominate the latency highly depends on the photodetector design.  State-of-the-art commercial photodetectors have up to 100 GHz bandwidth, thus contributing $\sim 10$ ps latency.  The value of $\tau_{RF}$ is variable and will depend on the use-case; in some scenarios the RF bandpass filter and amplifier are optional.  If using a narrow-band RF filter to remove spurious frequencies, then $\tau_{RF}$ will strongly dominate the physical latency.  Thus, one benefit of keeping the spurious frequencies is to reduce the latency.  Finally, the propagation time $\tau_{\text{prop}}$ is determined by the length of the optical and electrical paths.  The frequency-encoded electromagnetic waves will pass through these paths at approximately the speed of light, depending on the refractive index and waveguide properties.  The combined length of commercial fiber-optical components typically have tens of centimeters of optical path length after trimming the fiber leads, contributing $\sim 300$ ps of latency.  The electrical RF connections will contribute a similar latency.  The optical path length can be shortened to tens of millimeters by implementing this architecture on a photonic integrated circuit \cite{streshinsky2014silicon}, reducing the latency to $\sim 30$ ps.  Therefore depending on the scenario, the latency of this architecture will be dominated by data movement at the speed of light, $\tau_{\text{prop}}$.

In our experiment, we measured a latency of 60 ns using DPMZMs with 30 GHz bandwidth, a balanced photodetector with 45 MHz bandwidth, and an RF amplifier with 1 GHz bandwidth.  In addition, our experimental setup contains approximately 10 meters of signal propagation, given that standard commercial fiber components have 1 meter fiber leads on each side, plus the length of the RF coaxial cables.  Thus, in our experiment, the dominant sources of latency are $\tau_{PD} \approx 1/45\ \text{MHz} \approx 25\ \text{ns}$ and $\tau_{\text{prop}} \approx \frac{10\ \text{m}}{3\cdot 10^8 \text{m/s}}  \approx 35\ \text{ns}$.

For the RF filter analysis in (main text) Figure 5(c), because no electronic RF filters are required to implement a simple frequency-domain filter with MAFT, $\tau_{\text{RF}}=0$.  The final terms $\tau_{\text{Gabor}}$ and $\tau_{\text{JESD}}$ are included so that the analog MAFT-filtered signal can continue in the digital communications pipeline.  The latency contribution from $\tau_{\text{Gabor}}$ will be dominated by the Gabor limit regardless of whether an ADC or analog filter bank is used to read out the output signal.

\noindent \textbf{Gabor Limit:}  The fundamental limit to the latency of digitally reading out the signal is the amount of time it takes to read out the frequency mode: $\tau_{\text{Gabor}} = 1 / \Delta f$.

\noindent \textbf{Latency Specs:}  This latency estimation model takes common specifications from data sheets to estimate the latency.  Tables \ref{table:gpu} to \ref{table:maft} show the various parameters used to create the latency comparison plot in (main text) Figure 5(c).  

\noindent \textbf{Latency Breakdown:}  Figure \ref{fig:bdown} shows how much each variable contributes to the total latency of each evaluated RF receiver architecture.  As is already well-known, this illustrates that the primary bottleneck in electronic processing latency is data movement.  Hence, the Shannon capacity-limited data movement to compute is what gives MAFT the advantage in latency compared to electronic versions.

\bibliography{sample} 